\documentstyle[amstex,amssymb,12pt]{article}
\begin{document}

\author{{\bf Fabio Cardone}$^{a,b}${\bf , Alessio Marrani}$^{c}${\bf \ and } \and
{\bf Roberto Mignani}$^{b-d}$ \\
$a$ Istituto per lo Studio dei Materiali Nanostrutturati\\
(ISMN-CNR)\\
Via dei Taurini, 19\\
00185 ROMA, Italy\\
$b$ I.N.D.A.M. - G.N.F.M.\\
$c$ Dipartimento di Fisica ''E. Amaldi''\\
Universit\`{a} degli Studi ''Roma Tre''\\
Via della Vasca Navale, 84\\
00146 ROMA, Italy\\
$d$ I.N.F.N. - Sezione di Roma III}
\title{{\bf A new pseudo-Kaluza-Klein scheme for geometrical description of
interactions}}
\maketitle

\begin{abstract}
We illustrate the main features of a new Kaluza-Klein-like scheme (Deformed
Relativity in five dimensions). It is based on a five-dimensional Riemannian
space in which the four-dimensional space-time metric is deformed (i.e. it
depends on the energy) and energy plays the role of the fifth dimension. We
review the solutions of the five-dimensional Einstein equations in vacuum
and the geodetic equations in some cases of physical relevance. The Killing
symmetries of the theory for the energy-dependent metrics corresponding to
the four fundamental interactions (electromagnetic, weak, strong and
gravitational) are discussed for the first time. Possible developments of
the formalism are also briefly outlined.
\end{abstract}

\section{\bf \ Introduction}

The problem of the ultimate geometrical structure of the physical world -
both at a large and a small scale - is an old-debated one. After Einstein,
the generally accepted view is that physical phenomena do occur in a
four-dimensional manifold, with three spatial and one time dimensions, and
that space-time possesses a {\em global }Riemannian structure, whereas it is
{\em locally}{\it \ }flat (i.e. endowed with a Minkowskian geometry).

However, as is well known, many attempts at generalizing the
four-dimensional Einsteinian picture have been made in this century, mainly
aimed at building up unified schemes of the fundamental interactions. Such
efforts can be roughly divided into two main groups. In the former, the
existence of further dimensions is assumed ([1]-[9]) (by preserving the
usual Einsteinian structure of the 4-d. spacetime), whereas in the latter
one hypothesizes [10] global and/or local four-dimensional geometries,
different from the Minkowskian or the Riemannian ones (mainly of the Finsler
type [11]). The most celebrated theory of the first type is due to Kaluza
[2] and Klein [3], who assumed a five-dimensional space-time, in order to
unify gravitation and electromagnetism in a single geometrical structure. In
their scheme, the coefficient of the fifth coordinate is constant, whereas
Jordan [4] and Thiry [5] considered it a general function of the space-time
coordinates. The Kaluza-Klein formalism was since then extended to even
higher dimensions, in order to achieve unification of all four fundamental
interactions, i.e. including weak and strong forces ([6]-[8]). Modern
generalizations [8] of the Kakuza-Klein scheme require a {\em minimum}
number of 11 dimensions in order to accommodate the Standard Model of
electroweak and strong interactions (let us recall that 11 is also the{\it \
}{\em maximum} number of dimensions required by supergravity theories [9]).

In the last decade, two of us (F.C. and R.M.) introduced a generalization of
Special Relativity, called {\it Deformed Special Relativity (DSR) }[12]. It
was{\it \ }essentially aimed, in origin, at dealing in a phenomenological
way with a possible breakdown of local Lorentz invariance (LLI). Actually
the experimental data of some physical processes, ruled by different
fundamental interactions, seem indeed to provide evidence for local
departures from the usual Minkowski metric [12]. They are: the lifetime of
the (weakly decaying) $K_{s}^{0}$ meson [13]; the Bose-Einstein correlation
in (strong) pion production [14]; the superluminal propagation of
electromagnetic waves in waveguides [15]. All such phenomena seemingly show
a (local) breakdown of Lorentz invariance and, therefore, an inadequacy of
the Minkowski metric in describing them, {\em at different energy scales and
for the three interactions involved} (electromagnetic, weak and strong). On
the contrary, they apparently admit of a consistent interpretation in terms
of a {\em deformed Minkowski space-time, with metric coefficients depending
on the energy of the process considered }[12]. Moreover, it can be shown
that also the experimental results on the slowing down of clocks in a
gravitational field [16] can be described in terms of a deformed
energy-dependent metric [12].

DSR is just a (four-dimensional) generalization of the (local) space-time
structure based on an energy-dependent deformation of the usual Minkowski
geometry. What's more, the corresponding deformed metrics obtained from the
experimental data provide an {\em effective dynamical description of the
interactions}{\it \ }ruling the phenomena considered ({\em at least at the
energy scale and in the energy range considered}). Then one realizes, for
all four interactions, the so-called {\it ``Solidarity Principle``}, between
space-time and interaction (so that the peculiar features of every
interaction determine --- locally --- its own space-time structure), that
--- following B. Finzi [17] --- can be stated as follows: {\it ``Space-time
is solid with interactions, so that their respective properties affect
mutually''}.

Moreover, it was shown that the deformed Minkowski space with
energy-dependent metric {\em admits a natural embedding in a
five-dimensional space-time, with energy as extra dimension }([18], [12]).
Namely, the four-dimensional, deformed, energy-dependent space-time is only
a manifestation (a ''shadow'', to use the famous word of Minkowski) of a
larger, five-dimensional space, in which energy plays the role of the fifth
dimension. The new formalism one gets in this way ({\it Deformed Relativity
in Five Dimensions, }DR5) is a {\em Kaluza-Klein-like one, the main points
of departure from a standard KK scheme being the deformation of the
Minkowski space-time and the use of energy as extra dimension }(this last
feature entails, among the others, that the DR5 formalism is {\em %
noncompactified }).

DR5 is therefore a generalization of Einstein's Relativity sharing both
features of a change of the 4-d. Minkowski metric and the presence of extra
dimensions, and it permits also to give new intriguing insights on basic
properties, such as mass, of elementary particles, relating them to
fundamental geometrical quantities ([33], [34]).

The purpose of the present paper is to illustrate the DR5 formalism and to
give new results on the isometries of the five-dimensional space of the
theory.

The paper is organized as follows. Sect. 2 contains a brief review of the
formalism of the four-dimensional deformed Minkowski space and gives the
explicit expressions of the deformed metrics obtained, for the fundamental
interactions, by the phenomenological analysis of the experimental data. In
Sect.s 3-5 we illustrate the main features of the DR5 scheme. Its
geometrical structure --- based on a five-dimensional space in which the
four-dimensional space-time $\Re _{5}$ is deformed and the energy $E$ plays
the role of fifth dimension --- is discussed in Sect. 3. In Sect. 4 we write
down the related five-dimensional Einstein equations in general (with all
five metric coefficients depending on {\it E}, and including an arbitrary
''cosmological constant'' $\Lambda (E)$), and solve them explicitly in two
special cases of physical relevance and for $\Lambda {\it \ }=\ 0$. The
solutions obtained, with their physical meaning, are discussed in Sect. 5.
In Sect. 6 we derive the isometries of $\Re _{5}$ for the four
phenomenological metrics by exploiting the Killing equations of DR5. Sect.7
contains a brief discussion of the geodetic equations. Concluding remarks
and possible further developments of the formalism are put forward in Sect.
8.

\section{\bf \ Deformed Special Relativity}

\subsection{\bf Deformed Minkowski space-time}

Let us briefly review the main features of the formalism of the
(four-dimensional) deformed Minkowski space [12].

If $M(x,g_{SR},R)${\it \ }is the usual Minkowski space of the standard
Special Relativity (SR) (where $x$ is a fixed Cartesian frame), endowed with
the metric tensor
\begin{equation}
g_{SR}=diag(1,-1,-1,-1),  \label{SSR4}
\end{equation}
the deformed Minkowski space $\tilde{M}(x,g_{DSR},R)$ is the same vector
space on the real field as $M$, with the same frame $x$, but with metric $%
g_{DSR}$ given by\footnote{%
In the following, we shall employ the notation ''ESC on'' (''ESC off'') to
mean that the Einstein sum convention on repeated indices is (is not) used.}
\begin{eqnarray}
g_{DSR}(E) &=&diag(b_{0}^{2}(E),-b_{1}^{2}(E),-b\text{$_{2}^{2}(E)\
,-b_{3}^{2}(E))=$}  \nonumber \\
&&  \nonumber \\
&&\stackrel{\text{{\footnotesize ESC off}}}{=}\delta _{\mu \nu }\left[
b_{0}^{2}(E)\delta _{\mu 0}-b_{1}^{2}(E)\delta _{\mu 1}-b_{2}^{2}(E)\delta
_{\mu 2}-b_{3}^{2}(E)\delta _{\mu 3}\right] ,  \label{DSR4}
\end{eqnarray}
where the metric coefficients $\left\{ b_{\mu }^{2}(E){\it \ }\right\} $
{\it (}$\mu \ =\ 0,1,2,3$)${\it \ }$are (dimensionless) positive functions
of the energy $E$ of the process considered\footnote{$E$ is to be understood
as the energy measured by the detectors via their electromagnetic
interaction in the usual Minkowski space.}: $b_{\mu }^{2}=b_{\mu }^{2}(E)$.
The generalized infinitesimal metric interval in $\tilde{M}$ reads therefore
\begin{eqnarray}
ds^{2} &=&b_{0}^{2}(E)c^{2}\left( dt\right) ^{2}-b_{1}^{2}(E)\left(
dx\right) ^{2}-b_{2}^{2}(E)\left( dy\right) ^{2}-b_{3}^{2}(E)\left(
dz\right) ^{2}=  \nonumber \\
&&  \nonumber \\
&=&g_{\mu \nu ,DSR}dx^{\mu }dx^{\nu }=dx\ast dx,  \label{scalarprod}
\end{eqnarray}
with $x^{\mu }=(x^{0},x^{1},x^{2},x^{3})=(ct,x,y,z),$ $c$ being the usual
light speed {\em in vacuo}. The last equality in (\ref{scalarprod}) defines
the scalar product $\ast $ in the deformed Minkowski space $\tilde{M}$. The
relativity theory based on $\tilde{M}$ is called {\em Deformed Special
Relativity }(DSR){\it \ }[12].

We want to stress that --- although uncommon --- the use of an
energy-dependent space-time metric is not new. Indeed, it can be traced back
to Einstein himself. In order to account for the modified rate of a clock in
presence of a gravitational field, Einstein first generalized the expression
of the special-relativistic interval with metric (\ref{SSR4}), by
introducing a {\em ''time curvature''} as follows:
\begin{equation}
ds^{2}=\left( 1+\frac{2\phi }{c^{2}}\right) c^{2}\left( dt\right)
^{2}-\left( dx\right) ^{2}-\left( dy\right) ^{2}-\left( dz\right) ^{2},
\label{Eins1}
\end{equation}
where $\phi $ is the Newtonian gravitational potential. In the present
scheme, \ the reason whereby one considers energy as the variable upon which
the metric coefficients depend is twofold. On one side, it has a
phenomenological basis in the fact that we want to exploit this formalism in
order to derive the deformed metrics corresponding to physical processes,
whose experimental data are just expressed in terms of the energy of the
process considered. On the other hand, one expects on physical grounds that
a possible deformation of the space-time to be intimately related to the
energy of the concerned phenomenon (in analogy to the gravitational case,
where space-time curvature is determined by the energy-matter distribution).

Let us recall that the metric (\ref{DSR4}) is supposed to hold {\em locally}%
, i.e. in the space-time region where the process occurs. Moreover, it is
supposed to play a {\em dynamical} role, thus providing a geometric
description of the interaction considered, especially as far as nonlocal,
nonpotential forces are concerned. In other words, each interaction produces
its own metric, formally expressed by the metric tensor $g_{DSR}$, but
realized via different choices of the set of parameters $b_{\mu }(E)$. We
refer the reader to Ref. [12] for a more detailed discussion.

It is also worth to notice that the space-time described by the interval (%
\ref{scalarprod}) actually has zero curvature, and therefore{\it \ }{\em it
is not} a ''true'' Riemannian space (whence the term ''deformation'' used to
describe such a situation). Therefore, on this respect, the geometrical
description of the fundamental interactions based on the metric (\ref{DSR4})
is different from that adopted in General Relativity to describe
gravitation. Moreover, for each interaction the corresponding metric reduces
to the Minkowskian one, $g_{\mu \nu ,SR}$ , for a suitable value of the
energy, $E_{0},$ characteristic of the interaction considered (see below).
But the energy of the process is fixed, and cannot be changed at will. Thus,
although it would be in principle possible to recover the Minkowski space by
a suitable change of coordinates (e.g. by a {\em rescaling}), this would
amount to a mere mathematical operation, devoid of any physical meaning. In
our five-dimensional vision, in fact, the physics of the interaction
considered lies in the curvature of the five-dimensional metric, which
depends on energy\footnote{%
On the contrary, the four-dimensional sections at $E=constant$ \ are {\em %
''mathematically flat''}, since they have (four-dimensional) zero curvature.}%
.

Inside the deformed space-time, a {\em maximal causal speed} $u$ can be
defined, whose role is analogous to that of the light speed in vacuum for
the usual Minkowski space-time. It can be shown that, for an isotropic
3-dimensional space ($b_{1}=b_{2}=b_{3}=b)$, its expression is
\begin{equation}
u=\frac{b_{0}}{b}c.  \label{u}
\end{equation}
This speed $u$ can be considered as the speed of the interaction ruling the
process described by the deformation of the metric. It is easily seen that
there may be maximal causal speeds which are {\em superluminal}, depending
on the interaction considered, because
\begin{equation}
u\gtreqqless c\Longleftrightarrow \frac{b_{0}}{b}\gtreqqless 1.  \label{5}
\end{equation}

Starting from the deformed space-time $\tilde{M}$, one can develop the
Deformed Special Relativity in a straightforward way. For instance, the
generalized Lorentz transformations, i.e. those transformations which
preserve the interval (\ref{scalarprod}), for an isotropic three-space and
for a boost, say, along the $x$-axis, read as follows [12]
\begin{equation}
\left\{
\begin{array}{lll}
x^{\prime } & = & \tilde{\gamma}(x-vt); \\
&  &  \\
y^{\prime } & = & y; \\
&  &  \\
z^{\prime } & = & z; \\
&  &  \\
t^{\prime } & = & \tilde{\gamma}\left( t-\tilde{\beta}^{2}\dfrac{x}{v}%
\right) ,
\end{array}
\right.  \label{6}
\end{equation}
where $v${\it \ }is the relative speed of the reference frames, and
\begin{equation}
\tilde{\beta}=\frac{v}{u};  \label{7}
\end{equation}
\begin{equation}
\tilde{\gamma}=\left( 1-\tilde{\beta}^{2}\right) ^{-1/2}.  \label{8}
\end{equation}

It must be carefully noted that, like the metric, the generalized Lorentz
transformations, too, depend on the energy (through the deformed rapidity
parameter $\tilde{\beta}$: see the expression (\ref{u}) of the maximal speed
$u$). This means that one gets different transformation laws for different
values of $E$, but still with the same functional dependence on the energy,
so that the invariance of the deformed interval (\ref{scalarprod}) is always
ensured (provided the process considered does always occur via the {\em same}
interaction).

From the knowledge of the generalized Lorentz transformations in the
deformed Minkowski space $\tilde{M}$, ${\it \ }$it is easy to derive the
main kinematical and dynamical laws valid in DSR. For this topic and further
features of DSR the interested reader is referred to Ref.s [12], [19], [31]
and [32].

\subsection{Description of interactions by energy-dependent metrics}

We want instead to review the results obtained for the deformed metrics,
describing the four fundamental interactions - electromagnetic, weak, strong
and gravitational - , from the phenomenological analysis of the experimental
data [12]. First of all, let us stress that, in all the cases considered,%
{\it \ }{\em one gets evidence for a departure of the space-time metric from
the Minkowskian one }(at least in the energy range examined).

The explicit functional form of the DSR metric (\ref{DSR4}) for the four
interactions is as follows.
\[
\]

1) {\bf Electromagnetic interaction}. The experiments considered are those
on the superluminal propagation of e.m. waves in conducting waveguides with
variable section (first observed at Cologne in 1992) [15]. The introduction,
in this framework, of a deformed Minkowski space is motivated by ascribing
the {\em superluminal} speed of the signals to some {\em nonlocal} e.m.
effect, inside the narrower part of the waveguide, which can be described in
terms of an effective deformation of space-time inside the barrier region
[20]. Since we are dealing with electromagnetic forces (which are usually
described by the Minkowskian metric), we can assume $b_{0}^{2}=1$ (this is
also justified by the fact that all the relevant deformed quantities depend
actually on the ratio $b/b_{0}$). Assuming moreover an isotropically
deformed three-space ($b_{1}$ = $b_{2}$ = $b_{3}$ = $b$) \footnote{%
Notice that the assumption of {\em spatial isotropy} for the electromagnetic
interaction in the waveguide propagation is only a matter of convenience,
since waveguide experiments do not provide any physical information on space
directions different from the propagation one (the axis of the waveguide).
An analogous consideration holds true for the weak case, too (see below).},
one gets [12]
\begin{equation}
g_{DSR,e.m.}(E)=diag\left(
1,-b_{e.m.}^{2}(E),-b_{e.m.}^{2}(E),-b_{e.m.}^{2}(E)\right) ;  \label{em1}
\end{equation}
\begin{eqnarray}
b_{e.m.}^{2}(E) &=&\left\{
\begin{array}{lll}
(E/E_{0,e.m.})^{1/3}, & 0<E\leq E_{0,e.m.} &  \\
&  &  \\
1, & E_{0,e.m.}<E &
\end{array}
\right. = \\
&&  \nonumber \\
&&  \nonumber \\
&=&1+\Theta (E_{0,e.m.}-E)\left[ \left( \frac{E}{E_{0,e.m.}}\right) ^{1/3}-1%
\right] ,E>0,
\end{eqnarray}
(where $\Theta (x)$ is the Heaviside theta function, stressing the piecewise
structure of the metric). The threshold energy {\it E}$_{0,e.m.}$ is the
energy value at which the metric parameters are constant, i.e. the metric
becomes Minkowskian. The fit to the experimental data yields

\begin{equation}
E_{0,e.m.}=\left( 4.5\pm 0.2\right) \mu eV\,.  \label{em2}
\end{equation}
Notice that the value obtained for $E_{0}$ is of the order of the energy
corresponding to the coherence length of a photon for radio-optical waves ($%
E_{coh}\simeq 1\mu eV)${\it .}
\[
\]

2) {\bf Weak interaction.} The experimental input was provided by the data
on the pure leptonic decay of the meson $K_{s}^{0}$, whose lifetime $\tau
{\it \ }$is known in a wide energy range ($30\div 350$ $GeV$) [13] (an
almost unique case). Use has been made of the deformed law of time dilation
as a function of the energy, which reads [12]
\begin{equation}
\tau =\frac{\tau _{0}}{\left[ 1-\left( \dfrac{b}{b_{0}}\right) ^{2}+\left(
\dfrac{b}{b_{0}}\right) ^{2}\left( \dfrac{m_{0}}{E}\right) ^{2}\right] ^{1/2}%
}.  \label{tau}
\end{equation}
As in the electromagnetic case, an isotropic three-space was assumed,
whereas the {\em isochrony} with the usual Minkowski metric (i.e. $%
b_{0}^{2}=1$) was {\em derived} by the fit of (\ref{tau}) to the
experimental data. The corresponding metric is therefore given by
\begin{equation}
g_{DSR,weak}(E)=diag\left(
1,-b_{weak}^{2}(E),-b_{weak}^{2}(E),-b_{weak}^{2}(E)\right) ;  \label{weak1}
\end{equation}
\begin{eqnarray}
b_{weak}^{2}(E) &=&\left\{
\begin{array}{lll}
(E/E_{0,weak})^{1/3}, & 0<E\leq E_{0,weak} &  \\
&  &  \\
1, & E_{0,weak}<E &
\end{array}
\right. = \\
&&  \nonumber \\
&&  \nonumber \\
&=&1+\Theta (E_{0,weak}-E)\left[ \left( \frac{E}{E_{0,weak}}\right) ^{1/3}-1%
\right] ,E>0,
\end{eqnarray}
with

\begin{equation}
\text{ }E_{0,weak}=\left( 80.4\pm 0.2\right) GeV.  \label{weak2}
\end{equation}
Two points are worth stressing. First, the value of $E_{0,weak}$--- i.e. the
energy value at which the weak metric becomes Minkowskian --- corresponds to
the mass of the {\it W}-boson, through which the $K_{s}^{0}$-decay occurs.
Moreover, the leptonic metric (\ref{weak1})-(\ref{weak2}) has the same form
of the electromagnetic metric (\ref{em1})-(\ref{em2}). Therefore, one
recovers, by the DSR formalism, the well-known result of the
Glashow-Weinberg-Salam model that, at the energy scale $E_{0,weak}$, the
weak and the electromagnetic interactions are mixed. We want also to notice
that, in both the electromagnetic and the weak case, the metric parameter
exhibits a {\em ''sub-Minkowskian'' }behavior, i.e. $b(E)$ approaches $1$
{\em from below} as energy increases.
\[
\]

3) {\bf Strong interaction. }The phenomenon considered is the so-called
Bose-Einstein (BE) effect in the strong production of identical bosons {\bf %
\ }in high-energy collisions, which consists in an enhancement of their
correlation probability [14]. The DSR formalism permits to derive a
generalized BE correlation function, depending on all the four metric
parameters $b_{\mu }(E)$ [12]. By using the experimental data on pion pair
production, obtained in 1984 by the UA1 group at CERN [21], one gets the
following expression of the strong metric for the two-pion BE phenomenon
[12]:
\begin{equation}
g_{DSR,strong}(E)=diag\left(
b_{strong}^{2}(E),-b_{1,strong}^{2}(E),-b_{2,strong}^{2}(E),-b_{strong}^{2}(E)\right) ;
\label{strong1}
\end{equation}
\begin{eqnarray}
b_{strong}^{2}(E) &=&\left\{
\begin{array}{lll}
1, & 0<E\leq E_{0,strong} &  \\
&  &  \\
(E/E_{0,strong})^{2}, & E_{0,strong}<E &
\end{array}
\right. =  \label{16} \\
&&  \nonumber \\
&&  \nonumber \\
&=&\smallskip 1+\Theta (E-E_{0,strong})\left[ \left( \frac{E}{E_{0,strong}}%
\right) ^{2}-1\right] ,E>0;
\end{eqnarray}
\begin{equation}
b_{1,strong}^{2}(E)=\left( \sqrt{2}/5\right) ^{2};  \label{17}
\end{equation}
\begin{equation}
b_{2,strong}^{2}=(2/5)^{2},  \label{18}
\end{equation}
with

\begin{equation}
E_{0,strong}=\left( 367.5\pm 0.4\right) GeV.  \label{strong2}
\end{equation}
The threshold energy $E_{0,strong}$ is still the value at which the metric
becomes Minkowskian. Let us stress that, in this case, contrarily to the
electromagnetic and the weak ones, {\em a deformation of the time coordinate
occurs; }moreover, {\em the three-space is anisotropic}, with two spatial
parameters constant (but different in value) and the third one variable with
energy in an {\em ''over-Minkowskian''} way. It is also worth to recall that
the strong metric parameters $b_{\mu }$ admit of a sensible physical
interpretation: the spatial parameters are (related to) the spatial sizes of
the interaction region (''fireball'') where pions are produced, whereas the
time parameter is essentially the mean life of the process. We refer the
reader to Ref. [12] for further details.
\[
\]

4) {\bf Gravitation.} It is possible to show that the gravitational
interaction, too (at least on a {\em local} scale, i.e. in a neighborhood of
Earth) can be described in terms of an energy-dependent metric, whose time
coefficient was derived by fitting the experimental results on the relative
rates of clocks at different heights in the gravitational field of Earth
[16]. No information can be derived from the experimental data about the
space parameters. Physical considerations --- for whose details the reader
is referred to Ref. [12] --- lead to assume a gravitational metric of the
same type of the strong one, i.e. {\em spatially anisotropic} and with one
spatial parameter (say, $b_{3}$) equal to the time one: $%
b_{0}(E)=b_{3}(E)=b(E)$. The energy-dependent gravitational metric has
therefore the form
\begin{equation}
g_{DSR,grav}(E)=diag\left(
b_{grav}^{2}(E),-b_{1,grav}^{2}(E),-b_{2,grav}^{2}(E),-b_{grav}^{2}(E)%
\right) ;  \label{grav1}
\end{equation}
\begin{eqnarray}
b_{grav.}^{2}(E) &=&\left\{
\begin{array}{lll}
1, &  & 0<E\leq E_{0,grav.} \\
&  &  \\
\frac{1}{4}(1+E/E_{0,grav.})^{2}, &  & E_{0,grav.}<E
\end{array}
\right. =  \label{20'} \\
&&  \nonumber \\
&&  \nonumber \\
&=&1+\Theta (E-E_{0,grav.})\left[ \frac{1}{4}\left( 1+\frac{E}{E_{0,grav.}}%
\right) ^{2}-1\right] ,E>0
\end{eqnarray}
(the coefficients $b_{1,grav}^{2}(E)$\ and $b_{2,grav}^{2}(E)$\ are
presently {\em undetermined} at phenomenological level), with
\begin{equation}
E_{0,grav}=\left( 20.2\pm 0.1\right) \mu eV.  \label{grav2}
\end{equation}

The gravitational metric (\ref{grav1})-(\ref{grav2}) is {\em over-Minkowskian%
}, asymptotically Minkowskian with decreasing energy, like the strong one.
Intriguingly enough, the value of the threshold energy for the gravitational
case $E_{0grav}$ is approximately of the same order of magnitude of the
thermal energy corresponding to the $2.7^{o}K$ cosmic background radiation
in the Universe.

Moreover, the comparison of the values of the threshold energies for the
four fundamental interactions yields
\begin{equation}
E_{0,e.m.}<E_{0,grav}<E_{0,weak}<E_{0,strong},
\end{equation}
i.e. an increasing arrangement of $E_{0}$ from the electromagnetic to the
strong interaction. Moreover
\begin{equation}
\frac{E_{0,grav}}{E_{0,e.m.}}=4.49\pm 0.02\text{ };\text{ }\frac{E_{0,strong}%
}{E_{0,weak}}=4.57\pm 0.01,
\end{equation}
namely
\begin{equation}
\frac{E_{0,grav}}{E_{0,e.m.}}\simeq \frac{E_{0,strong}}{E_{0,weak}},
\end{equation}
an intriguing result indeed.

\section{Five-dimensional relativity with energy as extra dimension}

It is easily seen, from the examination of the {\em phenomenological metrics}
considered in the previous Section, that, in the DSR formalism, energy does
play a {\em dual} role. Indeed, on one side, $E$ is to be considered as a
{\em dynamical}{\it \ }variable, because it specifies the dynamical behavior
of the process under consideration, and, via the metric coefficients, it
provides us with a {\em dynamical map} - in the energy range of interest -
of the interaction ruling the given process. On the other hand, it
represents {\em a parameter characteristic of the phenomenon considered }%
(and therefore, for a given process, it {\em cannot} be changed{\em \ }at
will, as already stressed in the previous Section). In other words, when
describing a given process, the deformed geometry of space-time (in the
interaction region where the process is occurring) is {\em ''frozen''} at
the situation described by those values of the metric coefficients
corresponding to the energy value of the process considered. Otherwise
speaking, from a geometrical point of view, {\em all goes on as if we were
actually working on ''slices'' (sections) of a five-dimensional space, in
which the fifth dimension is just represented by the energy. }In other
words, a fixed value{\it \ }of the energy determines the space-time
structure of the interaction region for the given process {\em at that given
energy}{\it .} In this respect, therefore, $E$ is to be regarded as{\em \ a
geometrical quantity}, intimately connected to the very geometrical
structure of the physical world itself. The simplest way of taking into
account such a double role of $E${\em \ is to assume that energy does in
fact represent an extra dimension} --- besides the space and the time
ones---, {\em namely, to embed the deformed Minkowski space-time }$\tilde{M}$%
{\em \ in a larger, five-dimensional space} $\Re _{5}$ \ [18].

Let us specify the metric structure of the five-dimensional Riemann space $%
\Re _{5}$. We assume that the {\em generalized metric interval} in $\Re _{5}$
is given by
\begin{eqnarray}
ds_{(5)}^{2} &\equiv &b_{0}^{2}(E)c^{2}\left( dt\right)
^{2}-b_{1}^{2}(E)\left( dx\right) ^{2}-b_{2}^{2}(E)\left( dy\right)
^{2}-b_{3}^{2}(E)\left( dz\right) ^{2}\pm f(E)\ell _{0}^{2}\left( dE\right)
^{2}=  \nonumber \\
&&  \nonumber \\
&=&g_{\mu \nu ,DSR}(E)dx^{\mu }dx^{\nu }\pm f(E)(dx^{5})^{2}\equiv  \nonumber
\\
&&  \nonumber \\
&\equiv &g_{AB,DR5}(E)dx^{A}dx^{B},  \label{DR5metric1}
\end{eqnarray}
where $A,B=0,1,2,3,5,$ $x^{5}\equiv \ell _{0}E$, with $\ell _{0\text{ }}$
being a {\em dimensionally-transposing} constant (because of in this
Riemannian framework it is worth to give $x^{5}$ the dimension of a length, $%
\ell _{0\text{ }}$ has physical dimension [length]$\times $[energy]$^{-1}$),
and $f(E)>0$. The coefficients $\left\{ b_{\mu }^{2}(E)\right\} $ are those
determining the deformation of the 4-d. DSR spacetime \footnote{%
Since the metric coefficients $b_{\mu }^{2}(x^{5})$ and $f(x^{5})$ are {\em %
dimensionless}, they actually do depend on the ratio $\frac{x^{5}}{x_{0}^{5}}
$, where
\[
x_{0}^{5}\equiv \ell _{0}E_{0}
\]
is a {\em fundamental length}, proportional (by the {\em %
dimensionally-transposing} constant $\ell _{0}$) to the {\em threshold energy%
} $E_{0}$, characteristic of the interaction considered:
\begin{gather*}
b_{\mu }^{2}(x^{5})\equiv b_{\mu }^{2}\left( \frac{x^{5}}{x_{0}^{5}}\right)
=b_{\mu }^{2}\left( \frac{E}{E_{0}}\right) , \\
\\
f(x^{5})\equiv f\left( \frac{x^{5}}{x_{0}^{5}}\right) =f\left( \frac{E}{E_{0}%
}\right) .
\end{gather*}
{\footnotesize \ }For simplicity's sake, in the following we will omit, but
always understand, the cumbersome, but more rigorous notations $b_{\mu
}^{2}\left( \frac{x^{5}}{x_{0}^{5}}\right) $ and $f\left( \frac{x^{5}}{%
x_{0}^{5}}\right) $.}. The five-dimensional metric tensor $g_{DR5}$ reads
therefore
\begin{equation}
g_{AB,DR5}(x^{5})=diag(b_{0}^{2}(x^{5}),-b_{1}^{2}(x^{5}),-b_{2}^{2}(x^{5}),-b_{3}^{2}(x^{5}),\pm f(x^{5})),
\label{DR5metric2}
\end{equation}
and it is a function of the energy: $g_{DR5}=g_{DR5}(E).$

Some remarks are in order. First, in analogy with the space-time metric
coefficients $b_{\mu }$, we assumed that also the fifth metric coefficient
depends only on the energy: $f=f(E).$ However, one might assume that the
energy coefficient is a function {\em also} of the space-time coordinates $%
x=(x^{0},x^{1},x^{2},x^{3})$, namely $f=f(x,E)$. At present, such a
possibility will be disregarded. Moreover, we leave open the issue of
considering $E$ as a {\em timelike} or a {\em spacelike} coordinate in $\Re
_{5}$ (the double sign in front of $f$ in Eq.s (\ref{DR5metric1}) and (\ref
{DR5metric2}). Actually, in the {\em standard Kaluza-Klein scheme}, the
fifth dimension must necessarily be spacelike, because the number of
timelike dimensions cannot exceed one, if one wants to avoid {\em causal
anomalies} [22]. But --- and this is just another point worth stressing ---
{\em this five-dimensional scheme is not a ''true'' Kaluza-Klein one, }due
to the fact that the four-dimensional space-time is endowed with the
deformed metric (\ref{DSR4}). It is therefore an open issue whether or not,
in such a framework, more timelike dimensions do give rise to causal
anomalies.

We shall refer to the theory based on metric (\ref{DR5metric2}) as {\em %
Deformed Relativity in Five Dimensions (DR5)}. This approach is a {\em %
Kaluza-Klein-like}{\it \ }(or {\em pseudo-Kaluza-Klein}){\em \ }one, since
the four-dimensional space-time is endowed with the deformed metric (\ref
{DSR4}) and now the extra parameter is a physically sensible dimension.
Thus, on the latter respect, such a formalism belongs to the class of {\em %
noncompactified} KK\ theories, which, at the present status of experimental
knowledge, cannot be ruled out (see second Ref. in [8]). In the DR5
framework, the (deformed) Minkowski space-time is recovered not by means of
a {\em compactification} procedure, but instead by a {\em dimensional
reduction}.

As to considering energy as a {\em dynamical variable}, the use of momentum
components as dynamical variables on the same foot of the space-time ones
can be traced back to Ingraham [6]. Moreover, Dirac [23], Hoyle and Narlikar
[24] and Canuto et al. [25] treated mass as a dynamical variable in the
context of scale-invariant theories of gravity.

On the above side, the DR5 formalism has some connection with the
interesting ''Space-Time-Mass'' (STM) theory , in which the fifth dimension
is the rest mass, proposed by Wesson [26] and studied in detail by a number
of Authors. In either formalism it is assumed that {\em all} metric
coefficients do in general depend on the fifth coordinate. Such a feature
distinguishes both models from true Kaluza-Klein theories. However, the DR5
approach differs from the STM model (as well as from similar ones [27]) at
least in the following main respects:

$(i)$ its physical motivations are based on the phenomenological analysis of
Sect. 2, and therefore are not merely speculative;

$(ii)$ the fact of assuming {\em energy} (which is a {\em true variable}),
and not rest mass (which instead is an {\em invariant}), as fifth dimension%
\footnote{%
In this respect, therefore, DR5 rensembles more the formalism by Ingraham
[6].};

$(iii)$ the {\em local} (and not {\em global}) nature of the
five-dimensional space, whereby the energy-dependent deformation of the
four-dimensional space-time is assumed to provide a geometrical description
of the interactions [12].

The space $\Re _{5}$ has the following {\em ''slicing property''}
\[
\left. \Re _{5}\right| _{dx^{5}=0\Leftrightarrow x^{5}=\overline{x^{5}}}=%
\widetilde{M}(\overline{x^{5}})=\left\{ \widetilde{M}(x^{5})\right\} _{x^{5}=%
\overline{x^{5}}}
\]
(where $\overline{x^{5}}$ is a fixed value of the fifth coordinate)\ or, at
the level of the metric tensor:
\begin{eqnarray*}
\left. g_{AB,DR5}(x^{5})\right| _{dx^{5}=0\Leftrightarrow x^{5}=\overline{%
x^{5}}\in R_{0}^{+}} &=&diag\left( b_{0}^{2}(\overline{x^{5}}),-b_{1}^{2}(%
\overline{x^{5}}),-b_{2}^{2}(\overline{x^{5}}),-b_{3}^{2}(\overline{x^{5}}%
),\pm f(\overline{x^{5}})\right) = \\
&& \\
&=&g_{AB,DSR}(\overline{x^{5}}).
\end{eqnarray*}

\section{Five-dimensional Einstein equations}

\subsection{Solving the vacuum Einstein equations in $\Re _{5}$}

The {\em vacuum Einstein equations} in the space $\Re _{5}$ are [18]
\begin{equation}
R_{AB}-\frac{1}{2}g_{AB,DR5}R=\Lambda g_{AB,DR5},  \label{DR5Einstein}
\end{equation}
where $R_{AB}$ and $R=R_{A}^{A}$ are the five-dimensional {\em Ricci tensor}
and {\em scalar (intrinsic) curvature}, respectively, and $\Lambda $ is the
{\em ''cosmological'' constant}, which may, in principle, depend on both the
energy $E$ and the space-time coordinates $x:$ $\Lambda =\Lambda (x,E).$ As
is well known from Riemannian differential geometry, the {\em Ricci tensor}
explicitly reads (ESC on)
\begin{equation}
R_{AB}=\partial _{I}\Gamma _{AB}^{I}-\partial _{B}\Gamma _{AI}^{I}+\Gamma
_{AB}^{I}\Gamma _{IK}^{K}-\Gamma _{AI}^{K}\Gamma _{BK}^{I},  \label{24}
\end{equation}
with the {\em second-kind Christoffel symbols} $\Gamma _{AB}^{I}$ =$\left\{
\begin{array}{c}
I \\
AB
\end{array}
\right\} $ given by
\begin{equation}
2\Gamma _{AB}^{I}=g_{DR5}^{IK}(\partial _{B}g_{KA,DR5}+\partial
_{A}g_{KB,DR5}-\partial _{K}g_{AB,DR5}).  \label{25}
\end{equation}

We want here to consider some special cases of the five-dimensional Einstein
equations, which --- on account of the discussion of Section 2 --- are of a
special physical relevance. They are: $(i)$ the case of {\em spatial isotropy%
}; and: $(ii)$ when all the metric coefficients are {\em powers} of the
energy.

In order to simplify the notation, we write the metric tensor (\ref
{DR5metric2}) in the form
\begin{equation}
g_{DR5}(E)=diag(a(E),-b(E),-c(E),-d(E),f(E)),  \label{DR5metric3}
\end{equation}
(with $a(E),$ $b(E),c(E),d(E)$ positive functions) and adopt units such that
$c=$(velocity of light)$=1=\ell _{0}$. As can be seen by comparing the 5-d.
metrics (\ref{DR5metric3}) and (\ref{DR5metric2}), here we also re-adsorb
the ''$\pm $'' in a redefinition of $f(E)$, which now may change in sign. As
can be easily understood by looking at ODEs' systems (\ref{iso}) and (\ref
{Pow3}) and performing the {\em general functional reflection} $%
f(E)\rightarrow -f(E)$, such a redefinition of $f(E)$ is completely
uninfluential (in the sense that it leaves the results unchanged), {\em at
least} in the considered case of resolution of 5-d. Einstein equations (in
the reductive-simplifying cases {\bf i} and {\bf ii}) with {\em vacuum
prescription} (i.e. with $\Lambda =0$).
\[
\]

We have therefore:

{\bf Case i) - }For a {\em spatial isotropic deformation}, it is $%
b(E)=c(E)=d(E)$, so that the metric becomes
\begin{equation}
g_{DR5}(E)=diag(a(E),-b(E),-b(E),-b(E),f(E)).  \label{27}
\end{equation}
The independent Einstein equations obviously reduce to the following three
ones (henceforth, a prime denotes derivation with respect to $E$; moreover,
for simplicity of notation, we omit the explicit functional dependence of
all quantities on $E$):
\begin{equation}
\left\{
\begin{array}{l}
3(-2b^{\prime \prime }f+b^{\prime }f^{\prime })=4\Lambda bf^{2}; \\
\\
f\left[ a^{2}(b^{\prime })^{2}-2aa^{\prime }bb^{\prime }-4a^{2}bb^{\prime
\prime }-2aa^{\prime \prime }b^{2}+b^{2}(a^{\prime })^{2}\right] + \\
+abf^{\prime }(2ab^{\prime }+a^{\prime }b)=4\Lambda a^{2}b^{2}f^{2}; \\
\\
3b^{\prime }(ab)^{\prime }=-4\Lambda ab^{2}f.
\end{array}
\right.  \label{iso}
\end{equation}
\[
\]

{\bf Case ii) - }Since the space-time metric coefficients are {\em %
dimensionless}, as already pointed ou in Footnote 5, it is assumed that they
are functions of the ratio $E/E_{0}$ , where $E_{0\text{ }}$ is an energy
scale characteristic of the interaction (and the process) considered (for
instance, the energy threshold in the {\em phenomenological} metrics (\ref
{em1})-(\ref{grav2})). Precisely, for the metric $g_{DR5}$ written in the
form (\ref{DR5metric3}), we put ({\em ''Power Ansatz''})
\begin{equation}
\left\{
\begin{array}{ccc}
a(E) & = &
\begin{array}{c}
\begin{array}{c}
(E/E_{0})^{q_{0}}\text{ };
\end{array}
\end{array}
\\
b(E) & = &
\begin{array}{c}
\begin{array}{c}
(E/E_{0})^{q_{1}}\text{ };
\end{array}
\end{array}
\\
c(E) & = &
\begin{array}{c}
\begin{array}{c}
(E/E_{0})^{q_{2}}\text{ };
\end{array}
\end{array}
\\
d(E) & = & (E/E_{0})^{q_{3}}\text{ }
\end{array}
\right.  \label{Pow1}
\end{equation}
($q_{0},q_{1},q_{2},q_{3}\in R$). For the fifth metric coefficient $f(E)$ we
also assume
\begin{equation}
f(E)=(E/E_{0})^{r},r\in R,  \label{Pow2}
\end{equation}
being understood, as before, that $E_{0}=x_{0}^{5}/\ell _{0}=x_{0}^{5}$ in
the assumed units, where $\ell _{0}=1$. Of course, the Einstein equations
reduce now to the following algebraic equations in the five exponents $q_{0}$%
{\it , }$q_{1}${\it , }$q_{2}${\it , }$q_{3}${\it , }$r${\it :}
\begin{gather}  \label{Pow3}
\left\{
\begin{array}{l}
(2+r)(q_{3}+q_{1}+q_{2})-q_{1}^{2}-q_{2}^{2}-q_{3}^{2}-q_{1}q_{2}-q_{1}q_{3}-q_{2}q_{3}=4\Lambda (E/E_{0})^{r+2}%
\text{ }; \\
\\
(2+r)(q_{3}+q_{0}+q_{2})-q_{2}^{2}-q_{3}^{2}-q_{0}^{2}-q_{2}q_{3}-q_{2}q_{0}-q_{3}q_{0}=4\Lambda (E/E_{0})^{r+2}%
\text{ }; \\
\\
(2+r)(q_{3}+q_{0}+q_{1})-q_{1}^{2}-q_{3}^{2}-q_{0}^{2}-q_{1}q_{3}-q_{1}q_{0}-q_{3}q_{0}=4\Lambda (E/E_{0})^{r+2}%
\text{ }; \\
\\
(2+r)(q_{0}+q_{1}+q_{2})-q_{1}^{2}-q_{2}^{2}-q_{0}^{2}-q_{1}q_{2}-q_{1}q_{0}-q_{2}q_{0}=4\Lambda (E/E_{0})^{r+2}%
\text{ }; \\
\\
q_{1}q_{2}+q_{1}q_{3}+q_{1}q_{0}+q_{2}q_{3}+q_{2}q_{0}+q_{3}q_{0}=-4\Lambda
(E/E_{0})^{r+2}\text{ }.
\end{array}
\right.  \nonumber  \label{power} \\
\end{gather}
Of course, for consistency one has to impose the {\em compatibility condition%
} that $\Lambda $, too, is a power of the energy, and precisely one should
assume the following functional dependence:
\[
\Lambda (E/E_{0})\backsim (E/E_{0})^{-(r+2)};
\]
needless to say, the {\em vacuum prescription} $\Lambda =0$ is compatible
with this hypothesis.
\[
\]

Solving Einstein's equations in the five-dimensional, deformed space $\Re
_{5}$ in the general case is quite an impossible task. On the contrary, it
is possible to show [18] that, in the two special cases considered above,
some classes of solutions can be found for Eq.s (\ref{iso}) and (\ref{power}%
) (respectively corresponding to {\em spatial isotropy} and metric
coefficients which are {\em powers} of the energy), at least for $\Lambda =0$
. Notice that assuming a vanishing cosmological constant has the physical
motivation (at least as far as gravitation is concerned and one is not
interested into quantum effects) that $\Lambda $ is related to the vacuum
energy; experimental evidence shows that $\Lambda \simeq 3\cdot
10^{-52}m^{-2}$.

We recall moreover that Eq.s (\ref{DR5Einstein}) imply $R=-\frac{10}{3}%
\Lambda $. Being $\Lambda =0$ (and consequently $R=0$) the spaces we will
find are obviously {\em Ricci flat}. However, they differ, in general, from
a 5-dimensional flat space, as it can be easily checked by showing
explicitly that some components of the Riemann curvature tensor do {\em not}
vanish.
\[
\]

{\bf i) }In the former case ({\em spatial isotropy}), by putting $\Lambda =0$%
, the system of ordinary differential equations (\ref{iso}) takes the form
\begin{equation}
\left\{
\begin{array}{l}
-2b^{\prime \prime }f+b^{\prime }f^{\prime }=0\text{ }; \\
\\
f\left[ a^{2}(b^{\prime })^{2}-2aa^{\prime }bb^{\prime }-4a^{2}bb^{\prime
\prime }-2aa^{\prime \prime }b^{2}+b^{2}(a^{\prime })^{2}\right] + \\
+abf^{\prime }(2ab^{\prime }+a^{\prime }b)=0\text{ }; \\
\\
b^{\prime }(ab)^{\prime }=0\text{ }.
\end{array}
\right.  \label{iso2}
\end{equation}

If $a=const.$ (i.e. $a^{\prime }=0$), then the third equation of (\ref{iso2}%
) implies $b^{\prime }=0$ ; it is thence easy to see that the remaining
equations are identically satisfied. Hence the system (39) admits only the
solution $b=const.$, $f(E)$ undetermined, which can be shown to correspond
(modulo {\em rescaling}) to a {\em flat} 5-dimensional space. This entails,
as one should suspect, that a 5-dimensional Minkowski space can be a
solution of our system.

If $a$ is {\em not} a constant, then the third equation implies either $%
(i.1) $ $b^{\prime }=0,(ab)^{\prime }\neq 0$ or $(i.2)$ $b^{\prime }\neq
0,(ab)^{\prime }=0$.

Let us consider these two cases.\bigskip

{\bf (i.1)} In this case $b=const.$ and the system (\ref{iso2}) admits
solutions with $a(E)$ arbitrary and $f(E)$ determined by the only remaining
non-trivial equation, namely:
\begin{equation}
f[(a^{\prime })^{2}-2aa^{\prime \prime }]=-aa^{\prime }f^{\prime }.
\label{31''}
\end{equation}
Putting
\begin{equation}
A(E)=\frac{2aa^{\prime \prime }-(a^{\prime })^{2}}{aa^{\prime }}=\frac{%
f^{\prime }}{f},  \label{eq1}
\end{equation}
we get then
\begin{equation}
f(E)=ke^{\int^{E}A(\xi )d\xi }  \label{31''''}
\end{equation}
where $k$ is an integration constant. We remark that, if $f(E)=const.$, Eq. (%
\ref{eq1}) becomes
\begin{equation}
(a^{\prime })^{2}-2aa^{\prime \prime }=0.  \label{31'''''}
\end{equation}
It is easy to see that this equation admits the only solution
\begin{equation}
a(E)=\left( 1+\frac{E}{E_{0}}\right) ^{2},  \label{31''''''}
\end{equation}
with $E_{0}$ constant. Therefore, this shows that the gravitational metric (%
\ref{grav1}) corresponds to $f=const.$ , in the case of {\em spatial isotropy%
}.\bigskip

{\bf (i.2) }In this second case{\bf ,} it is not difficult to get the
following class of solutions:
\begin{equation}
\begin{array}{c}
f(E)=k\left[ b^{\prime }(E)\right] ^{2}\text{ }; \\
\\
a(E)=b(E)^{-1}\text{ },
\end{array}
\label{32}
\end{equation}
where $k$ is a {\em constant} (which fixes the sign of $f$) and $b(E)$ is an
arbitrary function of $E$.
\[
\]

{\bf ii) }Let us now discuss the case of the metric coefficients which are
{\em pure powers} of the energy. For $\Lambda =0$ Eq.s (\ref{power}) admit
of twelve possible classes of solutions, which can be classified according
to the values of the five-dimensional vector ${\bf \alpha \equiv }%
(q_{0},q_{1},q_{2},q_{3},r)\in R^{5}$ built up from the energy exponents of
the metric coefficients (see Eq.s (\ref{Pow1}) and (\ref{Pow2})). Explicitly
one has [18]
\begin{multline*}
\text{-{\bf \ Class (I):}} \\
{\bf \ } \\
{\bf \alpha }_{I}=\left( q_{2},-q_{2}\left( \dfrac{2q_{3}+q_{2}}{2q_{2}+q_{3}%
}\right) ,q_{2},q_{3},\dfrac{q_{3}^{2}-2q_{3}+2q_{2}q_{3}-4q_{2}+3q_{2}^{2}}{%
2q_{2}+q_{3}}\right) ; \\
\end{multline*}

- {\bf Class (II): }${\bf \alpha }_{II}=\left( 0,q_{1},0,0,q_{1}-2\right)
;\bigskip $

{\bf \ }

- {\bf Class (III): }${\bf \alpha }_{III}=\left(
q_{2},-q_{2},q_{2},q_{2},-2(1-q_{2})\right) ;\bigskip $

{\bf \ }

- {\bf Class (IV): }${\bf \alpha }_{IV}=\left( 0,0,0,q_{3},q_{3}-2\right)
;\bigskip ${\bf \ }

{\bf - Class (V): }${\bf \alpha }_{V}=\left(
-q_{3},-q_{3},-q_{3},q_{3},-(1+q_{3})\right) ;\bigskip ${\bf \ }

{\bf - Class (VI): }${\bf \alpha }_{VI}=\left( q_{0},0,0,0,q_{0}-2\right)
;\bigskip ${\bf \ }

{\bf - Class (VII): }${\bf \alpha }_{VII}=\left(
q_{0},-q_{0},-q_{0},-q_{0},-2-q_{0}\right) ;\bigskip $

{\bf - Class (VIII): }${\bf \alpha }_{VIII}=\left( 0,0,0,0,r\right)
;\bigskip $

{\bf \ }

{\bf - Class (IX): }${\bf \alpha }_{IX}=\left( 0,0,q_{2},0,-2+q_{2}\right)
;\medskip $

{\bf - Class (X)}:
\[
{\bf \alpha }_{X}=\left( q_{0},-\dfrac{q_{3}q_{0}+q_{2}q_{3}+q_{2}q_{0}}{%
q_{2}+q_{3}+q_{0}},q_{2},q_{3},r_{X}\right) {\bf \ ,}
\]

{\bf \ }

{\bf \ }with
\begin{gather*}
r_{X}=\dfrac{%
q_{3}^{2}+q_{3}q_{0}-2q_{3}+q_{2}q_{3}-2q_{2}+q_{2}q_{0}+q_{2}^{2}-2q_{0}+q_{0}^{2}%
}{q_{2}+q_{3}+q_{0}}; \\
\end{gather*}
\begin{multline*}
\text{{\bf - Class (XI):}} \\
{\bf \ } \\
{\bf \alpha }_{XI}=\left( q_{0},{\bf -}\dfrac{q_{2}(2q_{0}+q_{2})}{%
2q_{2}+q_{0}},q_{2},q_{2},\dfrac{%
3q_{2}^{2}-4q_{2}+2q_{2}q_{0}-2q_{0}+q_{0}^{2}}{2q_{2}+q_{0}}\right) ; \\
\end{multline*}

{\bf - Class (XII): }${\bf \alpha }_{XII}=\left( q_{0},q_{2},q_{2},{\bf -}%
\dfrac{q_{2}(2q_{0}+q_{2})}{2q_{2}+q_{0}},r_{XII}\right) $ , with
\begin{gather*}
\\
r_{XII}=\dfrac{%
q_{3}^{2}+q_{3}q_{0}-2q_{3}+q_{2}q_{3}-2q_{2}+q_{2}q_{0}+q_{2}^{2}-2q_{0}+q_{0}^{2}%
}{q_{2}+q_{3}+q_{0}}.
\end{gather*}

{\bf \ \ }In the following Subsection, we shall discuss the physical
relevance of the above solutions.

\subsection{Discussion of the solutions.}

As we said in the previous Subsection, in the case of {\em spatial isotropy}
the analytical solution of Eq. ((\ref{eq1})), for $f=const.$, yields
immediately the {\em gravitational metric} (\ref{grav1}).

On the other hand, the twelve classes of solutions found when assuming that
the metric coefficients are {\em powers} of the energy, allow one to
recover, as special cases, {\em all} the phenomenological metrics discussed
in Sect. 2 [18]. Let us write explicitly the {\em infinitesimal metric
interval} in $\Re _{5}$ in such a case:
\begin{gather}
ds_{(5)}^{2}=\left( \frac{E}{E_{0}}\right) ^{q_{0}}\left( dt\right)
^{2}-\left( \frac{E}{E_{0}}\right) ^{q_{1}}\left( dx\right) ^{2}-\left(
\frac{E}{E_{0}}\right) ^{q_{2}}\left( dy\right) ^{2}-\left( \frac{E}{E_{0}}%
\right) ^{q_{3}}\left( dz\right) ^{2}+\left( \frac{E}{E_{0}}\right)
^{r}\left( dE\right) ^{2}.  \nonumber \\
\end{gather}

Then, it is easily seen that the Minkowski metric is recovered from {\em all}
classes of solutions. Solution (VIII) corresponds directly to a Minkowskian
space-time, with the exponent $r$ of the fifth coefficient undetermined. In
the other cases, we have to put:

$q_{1}=0$ for class (II);

$q_{2}=0$ for classes (III) and (IX);

$q_{3}=0$ for (IV) and (V);

$q_{0}=0$ for (VI) and (VII)

(for all the previous solutions, it is $r=-2$);

$q_{2}=q_{3}=0$ for class (I);

$q_{2}=q_{3}=q_{0}=0$ for class (X);

$q_{2}=q_{0}=0$ for class (XI);

$q_{2}=q_{0}=0$ for class (XII).

The latter four solutions have $r=0$ , and therefore correspond to a
five-dimensional Minkowskian (and thus {\em flat}) space.

If we set:

$q_{1}=1/3$ in class (II);

$q_{3}=1/3$ in class (IV) or

$q_{2}=1/3$ in class (IX)

(corresponding in all three cases to the value $r=5/3$ for the exponent of
the fifth metric coefficient), we get a metric of the ''electroweak type''
(see Eq.s (\ref{em1})-(\ref{em2}), (\ref{weak1})-(\ref{weak2})), i.e. with
unit time coefficient and one space coefficient behaving as $(E/E_{0})^{1/3}$
, {\em but spatially anisotropic} , since two of the space metric
coefficients are constant and Minkowskian (precisely, the {\it y, z}
coefficients for class (II); the {\it x, y} coefficients for class (IV); and
the {\it x, z} ones for class (IX)). Notice that such an {\em anisotropy}
does {\em not} disagree with the phenomenological results; indeed, in the
analysis of the experimental data one was forced to assume spatial isotropy
in the electromagnetic and in the weak cases, simply because of the lack of
experimental information on two of the space dimensions.

Putting $q_{0}=1$ in class (VI), we find a metric which is spatially
Minkowskian, with a time coefficient linear in $E$, i.e. a (gravitational)
metric of the Einstein type (\ref{Eins1}).

Class (I) allows us to find as a special case a metric of the strong type
(see Eq.s (\ref{strong1})-(\ref{strong2})). This is achieved by setting $%
q_{2}=2$, whence we get
\[
q_{1}=-4(q_{3}+1)/(q_{3}+4);\text{ \ }r=(q_{3}^{2}+2q_{3}+4)/(q_{3}+4).
\]
Moreover, for $q_{3}=0$, it is $q_{1}=-1;r=1.$ In other words, we have a
solution corresponding to $a(E)=b(E)=(E/E_{0})^{2}$ and {\em spatially
anisotropic}, i.e. a metric of the type (\ref{strong1})-(\ref{strong2}).

Finally, the three classes (X)-(XII) admit as special case the gravitational
metric (\ref{grav1})-(\ref{grav2}), which is recovered by putting $q_{0}=2$
and $q_{1}=q_{2}=q_{3}=0$ (whence also $r=0$) and by a {\em rescaling} and a
{\em translation} of the energy parameter $E_{0}$.
\[
\]

In conclusion, we can state that {\em the formalism of DR5 permits to
recover, as solutions of the vacuum Einstein equations, all the
phenomenological energy-dependent metrics of the electromagnetic, weak,
strong and gravitational type}{\it \ }(and also the gravitational one of the
Einstein kind, Eq. (\ref{Eins1})).

\section{Killing symmetries in the space $\Re _{5}$.}

The topics concerning Deformed Relativity in 5 dimensions we expounded in
the previous Sections have been already discussed in literature [18]. In the
present Section, we shall deal for the first time with the problem of the
{\em metric automorphisms} (i.e. {\em isometries}) of the 5-d. Riemann space
$\Re _{5}$ of DR5 [28].

\subsection{General case.}

Let us discuss the {\em Killing symmetries} of the space $\Re _{5}$ [28].

The {\em Killing equations} for metric (\ref{DR5metric2}) read
\begin{equation}
\xi _{\lbrack A;B]}=0\Leftrightarrow \xi _{A;B}+\xi _{B;A}=0,
\end{equation}
where as usual $;A$ denotes Riemann covariant derivative with respect to $%
x^{A}$ and
\begin{equation}
\xi _{A}=\xi _{A}(x^{0},x^{1},x^{2},x^{3},x^{5})\equiv \xi _{A}(x^{B})
\end{equation}
is the covariant Killing 5-vector of $\Re _{5}$.

From the {\em Christoffel symbols} $\Gamma _{BC}^{A}$ of the metric $%
g_{AB,DR5}(x^{5})$ we get the following system of 15 coupled, partial
derivative differential equations (PDEs) in $\Re _{5}$ for the Killing
vector $\xi _{A}(x^{B})$:

\begin{gather}
\begin{array}{c}
f(x^{5})\xi _{0,0}(x^{A})\pm b_{0}(x^{5})b_{0}^{\prime }(x^{5})\xi
_{5}(x^{A})=0
\end{array}
;  \label{1} \\
\nonumber \\
\left.
\begin{array}{c}
\xi _{0,1}(x^{A})+\xi _{1,0}(x^{A})=0 \\
\xi _{0,2}(x^{A})+\xi _{2,0}(x^{A})=0 \\
\xi _{0,3}(x^{A})+\xi _{3,0}(x^{A})=0
\end{array}
\right\} \text{{\small \ type I conditions ;}} \\
\nonumber \\
\left.
\begin{array}{c}
b_{0}(x^{5})(\xi _{0,5}(x^{A})+\xi _{5,0}(x^{A}))-2b_{0}^{\prime }(x^{5})\xi
_{0}(x^{A})=0
\end{array}
\right\} \text{{\small type II condition; }} \\
\nonumber \\
\begin{array}{c}
f(x^{5})\xi _{1,1}(x^{A})\mp b_{1}(x^{5})b_{1}^{\prime }(x^{5})\xi
_{5}(x^{A})=0
\end{array}
; \\
\nonumber \\
\left.
\begin{array}{c}
\xi _{1,2}(x^{A})+\xi _{2,1}(x^{A})=0 \\
\xi _{1,3}(x^{A})+\xi _{3,1}(x^{A})=0
\end{array}
\right\} \text{{\small type I conditions; }} \\
\nonumber \\
\left.
\begin{array}{c}
b_{1}(x^{5})(\xi _{1,5}(x^{A})+\xi _{5,1}(x^{A}))-2b_{1}^{\prime }(x^{5})\xi
_{1}(x^{A})=0
\end{array}
\right\} \text{{\small type II condition; }} \\
\nonumber \\
\begin{array}{c}
f(x^{5})\xi _{2,2}(x^{A})\mp b_{2}(x^{5})b_{2}^{\prime }(x^{5})\xi
_{5}(x^{A})=0
\end{array}
; \\
\nonumber \\
\left.
\begin{array}{c}
\xi _{2,3}(x^{A})+\xi _{3,2}(x^{A})=0
\end{array}
\right\} \text{{\small type I condition;}} \\
\nonumber \\
\left.
\begin{array}{c}
b_{2}(x^{5})(\xi _{2,5}(x^{A})+\xi _{5,2}(x^{A}))-2b_{2}^{\prime }(x^{5})\xi
_{2}(x^{A})=0
\end{array}
\right\} \text{{\small type II condition; }} \\
\nonumber \\
\begin{array}{c}
f(x^{5})\xi _{3,3}(x^{A})\mp b_{3}(x^{5})b_{3}^{\prime }(x^{5})\xi
_{5}(x^{A})=0
\end{array}
; \\
\nonumber \\
\left.
\begin{array}{c}
b_{3}(x^{5})(\xi _{3,5}(x^{A})+\xi _{5,3}(x^{A}))-2b_{3}^{\prime }(x^{5})\xi
_{3}(x^{A})=0
\end{array}
\right\} \text{{\small type II condition; }} \\
\nonumber \\
\begin{array}{c}
2f(x^{5})\xi _{5,5}(x^{A})-f^{\prime }(x^{5})\xi _{5}(x^{A})=0.
\end{array}
\label{15}
\end{gather}

PDEs (\ref{1})-(\ref{15}) can be divided in ''fundamental'' equations and
''constraint'' equations (of type I and II). The above system is in general
{\em overdetermined}, i.e. its solutions will contain numerical coefficients
satisfying a given algebraic system. Its explicit solutions are given by
\begin{gather}
\xi _{\mu }(x^{A})=F_{\mu }(x^{A\neq \mu })+  \nonumber \\
\nonumber \\
\pm (-\delta _{\mu 0}+\delta _{\mu 1}+\delta _{\mu 2}+\delta _{\mu 3})b_{\mu
}(x^{5})b_{\mu }^{\prime }(x^{5})(f(x^{5}))^{-1/2}\int dx^{\mu
}F_{5}(x^{0},x^{1},x^{2},x^{3});
\end{gather}
\begin{equation}
\xi _{5}(x^{A})=(f(x^{5}))^{1/2}F_{5}(x^{0},x^{1},x^{2},x^{3}).
\end{equation}
The five unknown functions $F_{A}(x^{B\neq A})$ are restricted by the two
following types of conditions:
\begin{gather}
I)\text{ \ {\small Type I (Cardinality 4, }}\mu \neq \nu \neq \rho \neq
\sigma \text{{\small ):}}  \nonumber \\
\nonumber \\
\pm A_{\mu }(x^{5})G,_{\nu \rho \sigma }(x^{0},x^{1},x^{2},x^{3})+B_{\mu
}(x^{5})G,_{\mu \mu \nu \rho \sigma }(x^{0},x^{1},x^{2},x^{3})+  \nonumber \\
\nonumber \\
+b_{\mu }(x^{5})F_{\mu ,5}(x^{A\neq \mu })-2b_{\mu }^{\prime }(x^{5})F_{\mu
}(x^{A\neq \mu })=0;
\end{gather}
\begin{gather}
II)\text{ \ {\small Type II (Cardinality 6, symm. in }}{\small \mu ,\nu }%
\text{, }\mu \neq \nu \neq \rho \neq \sigma \text{{\small ):}}  \nonumber \\
\nonumber \\
\text{\ }F_{\mu ,\nu }(x^{A\neq \mu })+F_{\nu ,\mu }(x^{A\neq \nu })+
\nonumber \\
\nonumber \\
\pm (-\delta _{\mu 0}+\delta _{\mu 1}+\delta _{\mu 2}+\delta _{\mu 3})b_{\mu
}(x^{5})b_{\mu }^{\prime }(x^{5})(f(x^{5}))^{-1/2}G,_{\nu \nu \rho \sigma
}(x^{0},x^{1},x^{2},x^{3})+  \nonumber \\
\nonumber \\
\pm (-\delta _{\nu 0}+\delta _{\nu 1}+\delta _{\nu 2}+\delta _{\nu 3})b_{\nu
}(x^{5})b_{\nu }^{\prime }(x^{5})(f(x^{5}))^{-1/2}G,_{\mu \mu \rho \sigma
}(x^{0},x^{1},x^{2},x^{3})=0,
\end{gather}
where:
\begin{gather}
A_{\mu }(x^{5})\equiv (-\delta _{\mu 0}+\delta _{\mu 1}+\delta _{\mu
2}+\delta _{\mu 3})b_{\mu }(x^{5})(f(x^{5}))^{-1/2}\cdot  \nonumber \\
\nonumber \\
\cdot \left[ -\left( b_{\mu }^{\prime }(x^{5})\right) ^{2}+b_{\mu
}(x^{5})b_{\mu }^{\prime \prime }(x^{5})-\frac{1}{2}b_{\mu }(x^{5})b_{\mu
}^{\prime }(x^{5})f^{\prime }(x^{5})(f(x^{5}))^{-1}\right] ;  \label{A}
\end{gather}
\begin{equation}
B_{\mu }(x^{5})\equiv b_{\mu }(x^{5})(f(x^{5}))^{1/2};  \label{B}
\end{equation}
\begin{equation}
G(x^{0},x^{1},x^{2},x^{3})\equiv \int
dx^{0}dx^{1}dx^{2}dx^{3}F_{5}(x^{0},x^{1},x^{2},x^{3}).
\end{equation}

\subsection{The hypothesis $\Upsilon $ of functional independence.}

Let us consider the derivative with respect to $x^{\mu }$ of Type I
conditions (ESC\ off):
\begin{gather}
\partial _{\mu }I):\pm A_{\mu }(x^{5})G,_{\mu \nu \rho \sigma
}(x^{0},x^{1},x^{2},x^{3})+B_{\mu }(x^{5})G,_{\mu \mu \mu \nu \rho \sigma
}(x^{0},x^{1},x^{2},x^{3})=0\Leftrightarrow  \nonumber \\
\nonumber \\
\Leftrightarrow \pm A_{\mu }(x^{5})F_{5}(x^{0},x^{1},x^{2},x^{3})+B_{\mu
}(x^{5})F_{5,\mu \mu }(x^{0},x^{1},x^{2},x^{3})=0.
\end{gather}
If $G(x^{0},x^{1},x^{2},x^{3})$ satisfies the {\em Schwarz lemma} at {\em any%
} order, since $\mu \neq \nu \neq \rho \neq \sigma $, one gets
\begin{equation}
G,_{\mu \nu \rho \sigma
}(x^{0},x^{1},x^{2},x^{3})=G,_{0123}(x^{0},x^{1},x^{2},x^{3})(=F_{5}(x^{0},x^{1},x^{2},x^{3})),
\end{equation}
namely the function $F_{5}(x^{0},x^{1},x^{2},x^{3})$ is in $\frac{\partial }{%
\partial x^{\mu }}I)$ $\ \forall \mu =0,1,2,3$.

It is therefore{\em \ sufficient} to assume that at least a special index
\begin{equation}
\overline{\mu }\in \left\{ 0,1,2,3\right\} :\left\{
\begin{array}{l}
\nexists c_{\overline{\mu }}\in R_{0}:\pm A_{\overline{\mu }}(x^{5})=c_{%
\overline{\mu }}B_{\overline{\mu }}(x^{5})(\forall x^{5}\in R_{0}^{+}) \\
\\
A_{\overline{\mu }}(x^{5})\neq 0,B_{\overline{\mu }}(x^{5})\neq 0
\end{array}
\right.
\end{equation}
exists, such that $(\forall x^{0},x^{1},x^{2},x^{3}\in R)$
\begin{gather}
(F_{5}(x^{0},x^{1},x^{2},x^{3})=)G,_{0123}(x^{0},x^{1},x^{2},x^{3})=0=
\nonumber \\
\nonumber \\
=G,_{\overline{\mu }\overline{\mu }\overline{\mu }%
123}(x^{0},x^{1},x^{2},x^{3})(=F_{5,\overline{\mu }\overline{\mu }%
}(x^{0},x^{1},x^{2},x^{3}))
\begin{array}{c}
\Rightarrow \\
\stackrel{\text{{\footnotesize (in gen.)}}}{\nLeftarrow }
\end{array}
\nonumber \\
\nonumber \\
\begin{array}{c}
\Rightarrow \\
\stackrel{\text{{\footnotesize (in gen.)}}}{\nLeftarrow }
\end{array}
G,_{\mu \mu \mu 123}(x^{0},x^{1},x^{2},x^{3})(=F_{5,\mu \mu
}(x^{0},x^{1},x^{2},x^{3}))=0,\text{ }\forall \mu =0,1,2,3.
\end{gather}

In the following the {\em existence hypothesis}
\begin{eqnarray}
\exists \text{{\footnotesize \ (at least one) }}\overline{\mu } &\in
&\left\{ 0,1,2,3\right\} :\left\{
\begin{array}{l}
\nexists c_{\overline{\mu }}\in R_{0}:\pm A_{\overline{\mu }}(x^{5})=c_{%
\overline{\mu }}B_{\overline{\mu }}(x^{5})(\forall x^{5}\in R_{0}^{+}) \\
\\
A_{\overline{\mu }}(x^{5})\neq 0,B_{\overline{\mu }}(x^{5})\neq 0
\end{array}
\right.  \nonumber \\
&&  \label{HP}
\end{eqnarray}
will be called {\em ''}$\Upsilon ${\em \ hypothesis'' of functional
independence.}

\subsection{Solving Killing equations in $\Re _{5}$ in the hypothesis $%
\Upsilon $ of functional independence.}

In the hypothesis $\Upsilon $ of {\em functional independence} the
contravariant Killing 5-vector has the form (ESC off)
\begin{equation}
\xi _{A}(x^{B})=\left( b_{\mu }^{2}(x^{5})\widetilde{F_{\mu }}(x^{\nu \neq
\mu }),0\right)
\end{equation}
where the 4 unknown real functions of 3 real variables $\left\{ \widetilde{%
F_{\mu }}(x^{\rho \neq \mu })\right\} $ are solutions of the following
system of 6 (due to the symmetry in $\mu $ and $\nu $) non-linear PDEs:
\begin{equation}
b_{\mu }^{2}(x^{5})\frac{\partial \widetilde{F_{\mu }}(x^{\rho \neq \mu })}{%
\partial x^{\nu }}+b_{\nu }^{2}(x^{5})\frac{\partial \widetilde{F_{\nu }}%
(x^{\rho \neq \nu })}{\partial x^{\mu }}=0,\text{ \ }\mu ,\nu =0,1,2,3,(\mu
\neq \nu ),  \label{sys1}
\end{equation}
which is in general {\em overdetermined}.

Solving system (\ref{sys1}) yields the following expressions for the
components of the contravariant Killing 5-vector $\xi
^{A}(x^{0},x^{1},x^{2},x^{3},x^{5})$ satisfying the 15 Killing PDEs (\ref{1}%
)-(\ref{15}) in the {\em hypothesis }$\Upsilon ${\em \ of functional
independence} (\ref{HP}):
\begin{gather}
\xi ^{0}(x^{1},x^{2},x^{3})=\widetilde{F_{0}}(x^{1},x^{2},x^{3})=  \nonumber
\\
\nonumber \\
=d_{8}x^{1}x^{2}x^{3}+d_{7}x^{1}x^{2}+d_{6}x^{1}x^{3}+d_{4}x^{2}x^{3}+
\nonumber \\
\nonumber \\
+(d_{5}+a_{2})x^{1}+d_{3}x^{2}+d_{2}x^{3}+(a_{1}+d_{1}+K_{0});
\end{gather}
\begin{gather}
\xi ^{1}(x^{0},x^{2},x^{3})=-\widetilde{F_{1}}(x^{0},x^{2},x^{3})=  \nonumber
\\
\nonumber \\
=-h_{2}x^{0}x^{2}x^{3}-h_{1}x^{0}x^{2}-h_{8}x^{0}x^{3}-h_{4}x^{2}x^{3}-
\nonumber \\
\nonumber \\
-\left( h_{7}+e_{2}\right) x^{0}-h_{3}x^{2}-h_{6}x^{3}-\left(
K_{1}+h_{5}+e_{1}\right) ;
\end{gather}
\begin{gather}
\xi ^{2}(x^{0},x^{1},x^{3})=-\widetilde{F_{2}}(x^{0},x^{1},x^{3})=  \nonumber
\\
\nonumber \\
=-l_{2}x^{0}x^{1}x^{3}-l_{1}x^{0}x^{1}-l_{6}x^{0}x^{3}-l_{4}x^{1}x^{3}-
\nonumber \\
\nonumber \\
-\left( l_{5}+e_{4}\right) x^{0}-l_{3}x^{1}-l_{8}x^{3}-(l_{7}+K_{2}+e_{3});
\end{gather}
\begin{gather}
\xi ^{3}(x^{0},x^{1},x^{2})=-\widetilde{F_{3}}(x^{0},x^{1},x^{2})=  \nonumber
\\
\nonumber \\
=-m_{8}x^{0}x^{1}x^{2}-m_{7}x^{0}x^{1}-m_{6}x^{0}x^{2}-m_{4}x^{1}x^{2}-
\nonumber \\
\nonumber \\
-\left( m_{5}+g_{2}\right) x^{0}-m_{3}x^{1}-m_{2}x^{2}-(m_{1}+g_{1}+c);
\end{gather}
\begin{equation}
\xi ^{5}=0\neq \xi ^{5}(x^{0},x^{1},x^{2},x^{3},x^{5}).
\end{equation}
where (some of) the real parameters satisfy the algebraic system
\begin{gather}
\left\{
\begin{array}{l}
b_{0}^{2}(x^{5})\left[ d_{8}x^{2}x^{3}+d_{7}x^{2}+d_{6}x^{3}+\left(
d_{5}+a_{2}\right) \right] + \\
\\
+b_{1}^{2}(x^{5})\left[ h_{2}x^{2}x^{3}+h_{1}x^{2}+h_{8}x^{3}+\left(
h_{7}+e_{2}\right) \right] =0; \\
\\
\\
b_{0}^{2}(x^{5})\left( d_{8}x^{1}x^{3}+d_{7}x^{1}+d_{4}x^{3}+d_{3}\right) +
\\
\\
+b_{2}^{2}(x^{5})\left[ l_{2}x^{1}x^{3}+l_{1}x^{1}+l_{6}x^{3}+\left(
l_{5}+e_{4}\right) \right] =0; \\
\\
\\
b_{0}^{2}(x^{5})\left( d_{8}x^{1}x^{2}+d_{6}x^{1}+d_{4}x^{2}+d_{2}\right) +
\\
\\
+b_{3}^{2}(x^{5})\left[ m_{8}x^{1}x^{2}+m_{7}x^{1}+m_{6}x^{2}+\left(
m_{5}+g_{2}\right) \right] =0; \\
\\
\\
b_{1}^{2}(x^{5})\left( h_{2}x^{0}x^{3}+h_{1}x^{0}+h_{4}x^{3}+h_{3}\right) +
\\
\\
+b_{2}^{2}(x^{5})\left( l_{2}x^{0}x^{3}+l_{1}x^{0}+l_{4}x^{3}+l_{3}\right)
=0; \\
\\
\\
b_{1}^{2}(x^{5})\left( h_{2}x^{0}x^{2}+h_{8}x^{0}+h_{4}x^{2}+h_{6}\right) +
\\
\\
+b_{3}^{2}(x^{5})\left( m_{8}x^{0}x^{2}+m_{7}x^{0}+m_{4}x^{2}+m_{3}\right)
=0; \\
\\
\\
b_{2}^{2}(x^{5})\left( l_{2}x^{0}x^{1}+l_{6}x^{0}+l_{4}x^{1}+l_{8}\right) +
\\
\\
+b_{3}^{2}(x^{5})\left( m_{8}x^{0}x^{1}+m_{6}x^{0}+m_{4}x^{1}+m_{2}\right)
=0.
\end{array}
\right.  \nonumber \\
\end{gather}

\subsection{The ''Power Ansatz'' and the reductivity of the hypothesis $%
\Upsilon $ of functional independence.}

We want now to investigate if and when the simplifying $\Upsilon $
hypothesis (\ref{HP}) --- we exploited in order to solve the Killing
equations in $\Re _{5}$ --- is {\em reductive}. To this aim, one needs to
consider explicit forms of the 5-d. Riemannian metric $g_{AB,DR5}(x^{5})$.
As we have seen in Section 4, the {\em ''Power Ansatz''} allows one to
recover {\em all} the{\em \ phenomenological metrics} derived for the four
fundamental interactions. So it is worth considering such a case,
corresponding to a 5-d. metric of the form
\begin{gather}
g_{AB,DR5power}(x^{5})=  \nonumber \\
\nonumber \\
=diag\left( \left( \frac{x^{5}}{x_{0}^{5}}\right) ^{q_{0}},-\left( \frac{%
x^{5}}{x_{0}^{5}}\right) ^{q_{1}},-\left( \frac{x^{5}}{x_{0}^{5}}\right)
^{q_{2}},-\left( \frac{x^{5}}{x_{0}^{5}}\right) ^{q_{3}},\pm \left( \frac{%
x^{5}}{x_{0}^{5}}\right) ^{r}\right) \text{ \ },  \nonumber \\
\nonumber \\
q_{0},\text{\ }q_{1},q_{2},q_{3},r\in R,\text{ \ }A,B=0,1,2,3,5.
\label{DR5metricPow}
\end{gather}
Notice that, in comparison with Eq. (\ref{DR5metric3}) with {\em Ans\"{a}tze}
(\ref{Pow1}) and (\ref{Pow2}) implemented, here we re-extracted ''$\pm $''
from the fifth metric coefficient, whence $\left( \frac{x^{5}}{x_{0}^{5}}%
\right) ^{r}>0$ $\forall x^{5}\in R_{0}^{+}$.

From Eq.s (\ref{A}), (\ref{B}) and (\ref{DR5metricPow}) one thus gets:
\begin{eqnarray}
A_{\mu ,power}(x^{5}) &=&-\left( -\delta _{\mu 0}+\delta _{\mu 1}+\delta
_{\mu 2}+\delta _{\mu 3}\right) \frac{q_{\mu }}{2}\left( 1+\frac{r}{2}%
\right) \left( \frac{x^{5}}{x_{0}^{5}}\right) ^{\frac{3}{2}q_{\mu }-\frac{1}{%
2}r-2}=  \nonumber \\
&&  \nonumber \\
&=&A_{\mu ,power}(q_{\mu },r;x^{5}); \\
&&  \nonumber
\end{eqnarray}
\begin{equation}
B_{\mu ,power}(x^{5})=\left( \frac{x^{5}}{x_{0}^{5}}\right) ^{\frac{1}{2}%
q_{\mu }+\frac{1}{2}r}=B_{\mu ,power}(q_{\mu },r;x^{5}).
\end{equation}
Therefore:
\begin{equation}
\frac{\pm A_{\mu ,power}(q_{\mu },r;x^{5})}{B_{\mu ,power}(q_{\mu },r;x^{5})}%
=\pm \left( \delta _{\mu 0}-\delta _{\mu 1}-\delta _{\mu 2}-\delta _{\mu
3}\right) \frac{q_{\mu }}{2}\left( 1+\frac{r}{2}\right) \left( \frac{x^{5}}{%
x_{0}^{5}}\right) ^{q_{\mu }-r-2}.
\end{equation}
Since $x^{5}\in R_{0}^{+}$, one respectively gets:
\begin{eqnarray}
A_{\mu ,power}(q_{\mu },r;x^{5}) &\neq &0\Leftrightarrow \frac{q_{\mu }}{2}%
\left( 1+\frac{r}{2}\right) \neq 0\Leftrightarrow  \nonumber \\
&&  \nonumber \\
&\Leftrightarrow &\left\{
\begin{array}{c}
q_{\mu }\neq 0 \\
\\
1+\frac{r}{2}\neq 0\Leftrightarrow 2+r\neq 0
\end{array}
\right. ;
\end{eqnarray}
\begin{equation}
B_{\mu ,power}(q_{\mu },r;x^{5})\neq 0,\forall q_{\mu },r\in R.
\end{equation}
Therefore:
\begin{equation}
\frac{\pm A_{\mu ,power}(q_{\mu },r;x^{5})}{B_{\mu ,power}(q_{\mu },r;x^{5})}%
=c_{(\mu ;q_{\mu },r)}\in R_{(0)},\forall x^{5}\in R_{0}^{+}\Leftrightarrow
q_{\mu }-r-2=0.
\end{equation}

It follows that, if one assumes $A_{\mu ,power}(q_{\mu },r;x^{5})\neq 0$ and
$B_{\mu ,power}(q_{\mu },r;x^{5})\neq 0$, in the framework of the {\em %
''Power Ansatz''} for $g_{AB,DR5}(x^{5})$ the {\em hypothesis }$\Upsilon $%
{\em \ of functional independence} (\ref{HP}) becomes:
\begin{gather}
\exists \text{{\footnotesize \ (at least one) }}\overline{\mu }\in \left\{
0,1,2,3\right\} :\left\{
\begin{array}{c}
q_{\overline{\mu }}-\left( r+2\right) \neq 0 \\
\\
\left\{
\begin{array}{c}
q_{\overline{\mu }}\neq 0 \\
\\
r+2\neq 0
\end{array}
\right.
\end{array}
\right\} \Leftrightarrow  \nonumber \\
\nonumber \\
\Leftrightarrow q_{\overline{\mu }}\neq 0,r+2\neq 0,q_{\overline{\mu }}\neq
r+2.
\end{gather}

In other words, in the framework of the {\em ''Power Ansatz''} for the
metric tensor the reductive nature of the $\Upsilon $ hypothesis depends on
the value of the real parameters $q_{0},q_{1},q_{2},q_{3}$ and $r$,
exponents of the components of $g_{AB,DR5power}(x^{5})$.

The (here not explicitly considered) discussion of the possible reductivity
of the $\Upsilon $ hypothesis for the 12 classes of solutions of the 5-d.
Einstein equations in vacuum derived in Subsect. 4.1 (labelled by the 5-d.
real vector ${\bf \alpha \equiv }\left( q_{0},q_{1},q_{2},q_{3},r\right) $)
allows one to state that in 5 general cases such hypothesis of functional
independence is reductive indeed. The Killing equations can be explicitly
solved in such cases. We do not deal here with these general cases, and
confine ourselves to discussing the special cases of the 5-d.
phenomenological power metrics describing the four fundamental interactions
(see Subsect. 2.2).

\subsection{The phenomenological 5-d. metrics of fundamental interactions.}

Let us now consider the 4-d. metrics of the deformed Minkowski spaces $%
\widetilde{M}(x^{5})$ for the four fundamental interactions (e.m., weak,
strong and gravitational) (see Eq.s (\ref{em1})-(\ref{grav2})). In passing
from the deformed, special-relativistic 4-d. framework of DSR to the
general-relativistic 5-d. one of DR5 --- geometrically corresponding to the
{\em embedding} of the deformed 4-d. Minkowski spaces $\left\{ \widetilde{M}%
(x^{5})\right\} _{x^{5}\in R_{0}^{+}}$ (where $x^{5}$ is a constant,
non-metric parameter) in the 5-d. Riemann space $\Re _{5}$ (where $x^{5}$ is
a metric coordinate), in general the phenomenological metrics (\ref{em1})-(%
\ref{grav2}) take the following 5-d. form (as usual $A,B=0,1,2,3,5$, and $%
f(x^{5})\in R_{0}^{+}$ $\forall x^{5}\in R_{0}^{+}$):
\begin{gather}
g_{AB,DR5,e.m.}(x^{5})=  \nonumber \\
\nonumber \\
=diag\left( 1,-\left\{ 1+\Theta (x_{0,e.m.}^{5}-x^{5})\left[ \left( \frac{%
x^{5}}{x_{0,e.m.}^{5}}\right) ^{1/3}-1\right] \right\} ,\right.  \nonumber \\
\nonumber \\
-\left\{ 1+\Theta (x_{0,e.m.}^{5}-x^{5})\left[ \left( \frac{x^{5}}{%
x_{0,e.m.}^{5}}\right) ^{1/3}-1\right] \right\} ,  \nonumber \\
\nonumber \\
\left. -\left\{ 1+\Theta (x_{0,e.m.}^{5}-x^{5})\left[ \left( \frac{x^{5}}{%
x_{0,e.m.}^{5}}\right) ^{1/3}-1\right] \right\} ,\pm f(x^{5})\right) ;
\label{em5D}
\end{gather}
\begin{gather}
g_{AB,DR5,weak}(x^{5})=  \nonumber \\
\nonumber \\
=diag\left( 1,-\left\{ 1+\Theta (x_{0,weak}^{5}-x^{5})\left[ \left( \frac{%
x^{5}}{x_{0,weak}^{5}}\right) ^{1/3}-1\right] \right\} ,\right.  \nonumber \\
\nonumber \\
-\left\{ 1+\Theta (x_{0,weak}^{5}-x^{5})\left[ \left( \frac{x^{5}}{%
x_{0,weak}^{5}}\right) ^{1/3}-1\right] \right\} ,  \nonumber \\
\nonumber \\
\left. -\left\{ 1+\Theta (x_{0,weak}^{5}-x^{5})\left[ \left( \frac{x^{5}}{%
x_{0,weak}^{5}}\right) ^{1/3}-1\right] \right\} ,\pm f(x^{5})\right) ;
\label{weak5D}
\end{gather}
\begin{gather}
g_{AB,DR5,strong}(x^{5})=  \nonumber \\
\nonumber \\
=diag\left( 1+\Theta (x^{5}-x_{0,strong}^{5})\left[ \left( \frac{x^{5}}{%
x_{0,strong}^{5}}\right) ^{2}-1\right] ,-\left( \frac{\sqrt{2}}{5}\right)
^{2},\right.  \nonumber \\
\nonumber \\
\left. -\left( \frac{2}{5}\right) ^{2},-\left\{ 1+\Theta
(x^{5}-x_{0,strong}^{5})\left[ \left( \frac{x^{5}}{x_{0,strong}^{5}}\right)
^{2}-1\right] \right\} ,\pm f(x^{5})\right) ;  \label{strong5D}
\end{gather}
\begin{gather}  \label{grav5D}
g_{AB,DR5,grav.}(x^{5})=  \nonumber \\
\nonumber \\
=diag\left( 1+\Theta (x^{5}-x_{0,grav.}^{5})\left[ \frac{1}{4}\left( 1+\frac{%
x^{5}}{x_{0,grav.}^{5}}\right) ^{2}-1\right] ,-b_{1,grav.}^{2}(x^{5}),\right.
\nonumber \\
\nonumber \\
\left. -b_{2,grav.}^{2}(x^{5}),-\left\{ 1+\Theta (x^{5}-x_{0,grav.}^{5})
\left[ \frac{1}{4}\left( 1+\frac{x^{5}}{x_{0,grav.}^{5}}\right) ^{2}-1\right]
\right\} ,\pm f(x^{5})\right) .  \nonumber \\
\end{gather}

\subsection{Phenomenological 5-d.metrics and the reductivity of the
hypothesis $\Upsilon $ of functional independence.}

We want now to investigate the possible reductivity of the {\em hypothesis }$%
\Upsilon ${\em \ of functional independence} (\ref{HP}) for the 5-d. metrics
(\ref{em5D})-(\ref{grav5D}). Due to the {\em piecewise structure} of the
phenomenological metrics, we shall distinguish the two cases {\bf :} $%
0<x^{5}<x_{0}^{5}$ \ (case {\bf a}) and $x^{5}\geqslant x_{0}^{5}$ (case
{\bf b}).
\[
\]

{\bf I-II - Electromagnetic and weak interactions.}

{\bf Case a).}\ In this energy range the form of the metrics (\ref{em5D})
and (\ref{weak5D}) is:
\begin{equation}
g_{AB,DR5}(x^{5})=diag\left( 1,-\left( \frac{x^{5}}{x_{0}^{5}}\right)
^{1/3},-\left( \frac{x^{5}}{x_{0}^{5}}\right) ^{1/3},-\left( \frac{x^{5}}{%
x_{0}^{5}}\right) ^{1/3},\pm f(x^{5})\right) .
\end{equation}
Then, the $\Upsilon $ hypothesis (\ref{HP}) is not satisfied for $\mu =0$
but it does for $\mu =i=1,2,3$ under the following condition:
\begin{equation}
\frac{1}{x^{5}}+\frac{1}{2}\frac{f^{\prime }(x^{5})}{f(x^{5})}\neq
cf(x^{5})\left( x^{5}\right) ^{\frac{2}{3}},c\in R.  \label{constr}
\end{equation}
\medskip

By explicitly solving the corresponding Killing equations, one gets (under
constraint (\ref{constr})) the following general expression for the
contravariant Killing 5-vector $\xi ^{A}(x^{0},x^{1},x^{2},x^{3},x^{5})$ of
the 5-d. phenomenological electromagnetic and weak metrics ( $x^{5}\in
R_{0}^{+}$):
\begin{gather}
\xi ^{0}(x^{1},x^{2},x^{3},x^{5})=  \nonumber \\
\nonumber \\
=\Theta _{R}(x^{5}-x_{0}^{5})\left[ -\zeta ^{1}x^{1}-\zeta ^{2}x^{2}-\zeta
^{3}x^{3}+\zeta ^{5}\int dx^{5}f(x^{5})^{\frac{1}{2}}\right] +T^{0};
\end{gather}
\begin{gather}
\xi ^{1}(x^{0},x^{2},x^{3},x^{5})=  \nonumber \\
\nonumber \\
=\Theta _{R}(x^{5}-x_{0}^{5})\left[ -\zeta ^{1}x^{0}-\Sigma ^{1}\int
dx^{5}f(x^{5})^{\frac{1}{2}}\right] +\theta ^{3}x^{2}-\theta ^{2}x^{3}+T^{1};
\end{gather}
\begin{gather}
\xi ^{2}(x^{0},x^{1},x^{3},x^{5})=  \nonumber \\
\nonumber \\
=\Theta _{R}(x^{5}-x_{0}^{5})\left[ -\zeta ^{2}x^{0}-\Sigma ^{2}\int
dx^{5}f(x^{5})^{\frac{1}{2}}\right] -\theta ^{3}x^{1}+\theta ^{1}x^{3}+T^{2};
\end{gather}
\begin{gather}
\xi ^{3}(x^{0},x^{1},x^{2},x^{5})=  \nonumber \\
\nonumber \\
=\Theta _{R}(x^{5}-x_{0}^{5})\left[ -\zeta ^{3}x^{0}-\Sigma ^{3}\int
dx^{5}f(x^{5})^{\frac{1}{2}}\right] +\theta ^{2}x^{1}-\theta ^{1}x^{2}+T^{3};
\end{gather}
\begin{gather}
\xi ^{5}(x^{0},x^{1},x^{2},x^{3},x^{5})=  \nonumber \\
\nonumber \\
=\Theta _{R}(x^{5}-x_{0}^{5})\left\{ \mp f(x^{5})^{-\frac{1}{2}}[\zeta
^{5}x^{0}+\Sigma ^{1}x^{1}+\Sigma ^{2}x^{2}+\Sigma ^{3}x^{3}-T^{5}]\right\} ,
\end{gather}
where the parameters have been suitably redefined and we introduced the
distribution $\Theta _{R}(x^{5}-x_{0}^{5})$ ({\em right specification} of
the Heaviside distribution $\Theta (x^{5}-x_{0}^{5})$):
\begin{equation}
\Theta _{R}(x^{5}-x_{0}^{5})\equiv \left\{
\begin{array}{l}
1,\text{ }x^{5}\geqslant x_{0}^{5} \\
\\
0,0<x^{5}<x_{0}^{5}
\end{array}
\right. .
\end{equation}

Thus, in the energy range $x^{5}\geqslant x_{0}^{5}$ the Killing group of
the {\em ''slices''} at $dx^{5}=0$ of $\Re _{5}$ is the {\em standard
Poincar\'{e} group}
\begin{equation}
P(3,1)_{STD.}=SO(3,1)_{STD.}\otimes _{s}Tr.(3,1)_{STD.},
\end{equation}
whereas for $0<x^{5}<x_{0}^{5}$ the 5-d. Killing group is
\begin{equation}
SO(3)_{STD.}\otimes _{s}Tr.(3,1)_{STD.}.
\end{equation}
\medskip

In the considered energy range $0<x^{5}<x_{0}^{5}$ , if the hypothesis $%
\Upsilon $ of functional independence (\ref{HP}) does {\em not} hold for any
value of $\mu $, then (\ref{constr}) is violated, i.e. the metric
coefficient $f(x^{5})$ satisfies the following ordinary differential
equation (ODE):
\begin{equation}
\frac{1}{2}\frac{f^{\prime }(x^{5})}{f(x^{5})}-cf(x^{5})\left( x^{5}\right)
^{\frac{2}{3}}+\frac{1}{x^{5}}=0,c\in R.
\end{equation}
Such ODE belongs to the {\em homogeneous class of type G} and to the {\em %
special rational subclass of Bernoulli's ODEs} (it becomes {\em separable}
for $c=0$) . By solving it we get the following expression for the 5-d.
metric describing e.m. and weak interactions:
\begin{gather}
g_{AB,DR5}(x^{5})=diag\left( 1,-\left( \frac{x^{5}}{x_{0}^{5}}\right)
^{1/3},-\left( \frac{x^{5}}{x_{0}^{5}}\right) ^{1/3},-\left( \frac{x^{5}}{%
x_{0}^{5}}\right) ^{1/3},\right.  \nonumber \\
\nonumber \\
\left. \pm \left( 6c\left( \frac{x^{5}}{x_{0}^{5}}\right) ^{\frac{5}{3}%
}+\gamma \left( \frac{x^{5}}{x_{0}^{5}}\right) ^{2}\right) ^{-1}\right) ,
\end{gather}
with
\begin{equation}
c,\gamma \in R:6c\left( \frac{x^{5}}{x_{0}^{5}}\right) ^{\frac{5}{3}}+\gamma
\left( \frac{x^{5}}{x_{0}^{5}}\right) ^{2}>0,\forall x^{5}\in
R_{0}^{+}\Leftrightarrow c,\gamma \in R^{+}\text{ (not both zero),}
\end{equation}
valid in the energy range $0<x^{5}<x_{0}^{5}$ if the hypothesis $\Upsilon $ (%
\ref{HP}) is {\em not} satisfied.

Solving the relevant Killing equations yields the following expression for
the contravariant Killing 5-vector $\xi ^{A}(x^{0},x^{1},x^{2},x^{3},x^{5})$
corresponding to the e.m. and weak metrics:
\begin{equation}
\xi ^{0}=c_{0};
\end{equation}
\begin{equation}
\xi ^{1}(x^{2},x^{3})=-\left( a_{2}x^{2}+a_{3}x^{3}+a_{4}\right) \left(
x_{0}^{5}\right) ^{1/3};
\end{equation}
\begin{equation}
\xi ^{2}(x^{1},x^{3})=\left( a_{2}x^{1}b_{1}x^{3}-b_{6}\right) \left(
x_{0}^{5}\right) ^{1/3};
\end{equation}
\begin{equation}
\xi ^{3}(x^{1},x^{2})=\left( a_{3}x^{1}-b_{1}x^{2}-b_{2}\right) \left(
x_{0}^{5}\right) ^{1/3};
\end{equation}
\begin{equation}
\xi ^{5}=0.
\end{equation}
The 5-d. Killing group of isometries is therefore
\begin{equation}
SO(3)_{STD.\left( E_{3}\right) }\otimes _{s}Tr.(3,1)_{STD.}
\end{equation}
where $E_{3}$ is the 3-d. manifold with metric
\begin{equation}
g_{ij}=-\left( \frac{x^{5}}{x_{0}^{5}}\right) ^{1/3}diag\left( 1,1,1\right) .
\end{equation}
\bigskip\

{\bf \ Case b). }In this energy range the 5-d. metrics (\ref{em5D}) and (\ref
{weak5D}) read:
\begin{equation}
g_{AB,DR5}(x^{5})=diag\left( 1,-1,-1,-1,\pm f(x^{5})\right) .
\end{equation}
Therefore the hypothesis $\Upsilon $ of functional independence (\ref{HP})
is {\em not} satisfied $\forall \mu \in \left\{ 0,1,2,3\right\} $.
\[
\]

{\bf III - Strong interaction. }

{\bf Case a).} The metric (\ref{strong5D}) has the form
\begin{equation}
g_{AB,DR5}(x^{5})=diag\left( 1,-\frac{2}{25},-\frac{4}{25},-1,\pm
f(x^{5})\right) ,
\end{equation}
which does {\em not} satisfy the $\Upsilon $ -hypothesis (\ref{HP}) $\forall
\mu \in \left\{ 0,1,2,3\right\} $.\bigskip

C{\bf ase b)}. The 5-d. metric (\ref{strong5D}) reads:
\begin{equation}
g_{AB,DR5}(x^{5})=diag\left( \left( \frac{x^{5}}{x_{0}^{5}}\right) ^{2},-%
\frac{2}{25},-\frac{4}{25},-\left( \frac{x^{5}}{x_{0}^{5}}\right) ^{2},\pm
f(x^{5})\right) ,
\end{equation}
and the hypothesis $\Upsilon $ of functional independence (\ref{HP}) ({\em %
not} satisfied for $\mu =1,2$) holds true for $\mu =0,3$ under the
condition:
\begin{equation}
\frac{1}{x^{5}}+\frac{1}{2}\frac{f^{\prime }(x^{5})}{f(x^{5})}\neq c\frac{%
f(x^{5})}{x^{5}},c\in R.  \label{constr2}
\end{equation}
\medskip

By solving the relevant Killing equations under condition (\ref{constr2}),
one gets (after a suitable redenomination of the parameters) the following
general form of the contravariant Killing 5-vector $\xi
^{A}(x^{0},x^{1},x^{2},x^{3},x^{5})$ for the 5-d. phenomenological metric of
the strong interaction:
\begin{gather}
\xi ^{0}(x^{1},x^{2},x^{3},x^{5})=  \nonumber \\
\nonumber \\
=\Theta _{R}(x_{0}^{5}-x^{5})\left[ -\frac{2}{25}\zeta ^{1}x^{1}-\frac{4}{25}%
\zeta ^{2}x^{2}+\zeta ^{5}\int dx^{5}f(x^{5})^{\frac{1}{2}}\right] -\zeta
^{3}x^{3}+T^{0};
\end{gather}
\begin{gather}
\xi ^{1}(x^{0},x^{2},x^{3},x^{5})=  \nonumber \\
\nonumber \\
=\Theta _{R}(x_{0}^{5}-x^{5})\left[ -\zeta ^{1}x^{0}-\theta ^{2}x^{3}-\Sigma
^{1}\int dx^{5}f(x^{5})^{\frac{1}{2}}\right] +2\theta ^{3}x^{2}+T^{1};
\end{gather}
\begin{gather}
\xi ^{2}(x^{0},x^{1},x^{3},x^{5})=  \nonumber \\
\nonumber \\
=\Theta _{R}(x_{0}^{5}-x^{5})\left[ -\zeta ^{2}x^{0}+\theta ^{1}x^{3}-\Sigma
^{2}\int dx^{5}f(x^{5})^{\frac{1}{2}}\right] -\theta ^{3}x^{1}+T^{2};
\end{gather}
\begin{gather}
\xi ^{3}(x^{0},x^{1},x^{2},x^{5})=  \nonumber \\
\nonumber \\
=\Theta _{R}(x_{0}^{5}-x^{5})\left[ \frac{2}{25}\theta ^{2}x^{1}-\frac{4}{25}%
\theta ^{1}x^{2}-\Sigma ^{3}\int dx^{5}f(x^{5})^{\frac{1}{2}}\right] -\zeta
^{3}x^{0}+T^{3};
\end{gather}
\begin{gather}
\xi ^{5}(x^{0},x^{1},x^{2},x^{3},x^{5})=  \nonumber \\
\nonumber \\
=\Theta _{R}(x_{0}^{5}-x^{5})\left\{ \mp \left( f(x^{5})\right) ^{-\frac{1}{2%
}}[\zeta ^{5}x^{0}+\frac{2}{25}\Sigma ^{1}x^{1}+\frac{4}{25}\Sigma
^{2}x^{2}+\Sigma ^{3}x^{3}-T^{5}]\right\} .  \nonumber \\
\end{gather}

Thus, in the energy range $0<x^{5}\leqslant x_{0}^{5}$ the Killing group of
the {\em ''slices''} at $dx^{5}=0$ of $\Re _{5}$ is the {\em standard
Poincar\'{e} group} (suitably rescaled)
\begin{equation}
\left. \left[ P(3,1)_{STD.}=SO(3,1)_{STD.}\otimes _{s}Tr.(3,1)_{STD.}\right]
\right| _{x^{1}\longrightarrow \frac{\sqrt{2}}{5}x^{1},x^{2}\longrightarrow
\frac{2}{5}x^{2}},
\end{equation}
whereas for $x^{5}>x_{0}^{5}$ the 4-d Killing group is
\begin{equation}
\left( SO(2)_{STD.,\Pi (x^{1},x^{2}\longrightarrow \sqrt{2}x^{2})}\otimes
B_{x^{3},STD.}\right) \otimes _{s}Tr.(3,1)_{STD.}.
\end{equation}
Here
\begin{eqnarray*}
SO(2)_{STD.,\Pi (x^{1},x^{2}\longrightarrow \sqrt{2}x^{2})}
&=&SO(2)_{STD.,\Pi (x^{1}\longrightarrow \frac{\sqrt{2}}{5}%
x^{1},x^{2}\longrightarrow \frac{2}{5}x^{2})} \\
&&
\end{eqnarray*}
is the 1-parameter group (generated by the usual, special-relativistic
generator $\left. S_{SR}^{3}\right| _{x^{2}\longrightarrow \sqrt{2}x^{2}}$)
of the 2-d. rotations in the plane $\Pi (x^{1},x^{2})$ characterized by the
{\em coordinate contractions} $x^{1}\longrightarrow \frac{\sqrt{2}}{5}x^{1}$
, $x^{2}\longrightarrow \frac{2}{5}x^{2}$, and $B_{x^{3},STD.}$ is the usual
one-parameter group (generated by the special-relativistic generator $%
K_{SR}^{3}$) of the standard Lorentzian boosts along $\widehat{x^{3}}$%
.\medskip

In the energy range $x^{5}>x_{0}^{5}$ , when the hypothesis $\Upsilon $ of
functional independence (\ref{HP}) is {\em not} satisfied for any value of $%
\mu $ , the metric coefficient $f(x^{5})$ obeys the following equation:
\begin{equation}
\frac{1}{2}\frac{f^{\prime }(x^{5})}{f(x^{5})}-c\frac{f(x^{5})}{x^{5}}+\frac{%
1}{x^{5}}=0,c\in R.
\end{equation}
Such ODE is {\em separable} $\forall c\in R$. By solving it one gets the
following form of the 5-d. metric of the strong interaction (for $%
x^{5}>x_{0}^{5}$, and when the $\Upsilon $ \ hypothesis (\ref{HP}) is {\em %
not} satisfied):
\begin{gather}
g_{AB,DR5}(x^{5})=  \nonumber \\
\nonumber \\
=diag\left( \left( \frac{x^{5}}{x_{0}^{5}}\right) ^{2},-\frac{2}{25},-\frac{4%
}{25},-\left( \frac{x^{5}}{x_{0}^{5}}\right) ^{2},\pm \frac{1}{\gamma \left(
\frac{x^{5}}{x_{0}^{5}}\right) ^{2}+c}\right) ,
\end{gather}
with
\begin{equation}
c,\gamma \in R:\gamma \left( \frac{x^{5}}{x_{0}^{5}}\right) ^{2}+c>0,\forall
x^{5}\in R_{0}^{+}\Leftrightarrow c,\gamma \in R^{+}\text{ (not both zero).}
\end{equation}
Solving the related Killing equations yields the following contravariant
Killing 5-vector $\xi ^{A}(x^{0},x^{1},x^{2},x^{3},x^{5})$:
\begin{equation}
\xi ^{0}(x^{3};c,\gamma )=\left( 1-\delta _{c,0}\right) \left[ -\left(
x_{0}^{5}\right) ^{2}\left( \left( 1-\delta _{\gamma ,0}\right)
d_{3}x^{3}+T_{0}\right) \right] ;
\end{equation}
\begin{equation}
\xi ^{1}(x^{2};\gamma )=-\left( 1-\delta _{\gamma ,0}\right) \frac{25}{2}%
d_{2}x^{2}-\frac{25}{2}T_{1};
\end{equation}
\begin{equation}
\xi ^{2}(x^{1};\gamma )=\left( 1-\delta _{\gamma ,0}\right) \frac{25}{4}%
d_{2}x^{1}-\frac{25}{4}T_{2};
\end{equation}
\begin{equation}
\xi ^{3}(x^{0};c,\gamma )=-\left( 1-\delta _{c,0}\right) \left( 1-\delta
_{\gamma ,0}\right) \left( x_{0}^{5}\right) ^{2}\left(
d_{3}x^{0}+T_{3}\right) ;
\end{equation}
\begin{equation}
\xi ^{5}\left( x^{5};c,\gamma \right) =\pm \delta _{c,0}\frac{\gamma \alpha
}{\left( x_{0}^{5}\right) ^{2}}x^{5},
\end{equation}
where we evidenced the parametric dependence of $\xi ^{A}$ on $c$ and $%
\gamma $, and introduced the Kronecker $\delta $.

The 4-d. Killing group (i.e. of the slices at $dx^{5}=0$) is thus:
\begin{gather}
\left[ Tr._{\widehat{x^{1}},\widehat{x^{2}}\text{ \ }STD.}\otimes \left(
1-\delta _{c,0}\right) Tr._{\widehat{x^{0}}\text{ }STD.}\otimes \left(
1-\delta _{c,0}\right) \left( 1-\delta _{\gamma ,0}\right) Tr._{\widehat{%
x^{3}}\text{ }STD.}\right] \otimes _{s}  \nonumber \\
\nonumber \\
\otimes _{s}\left[ \left( 1-\delta _{\gamma ,0}\right) SO(2)_{STD.\left( \Pi
_{2}\right) }\otimes \left( 1-\delta _{c,0}\right) \left( 1-\delta _{\gamma
,0}\right) B_{STD.\text{ \ }\widehat{x^{3}}}\right] ,
\end{gather}
where $\Pi _{2}$ is the 2-d. manifold $\left( x_{1},x^{2}\right) $ with {\em %
''metric rescaling''} $x^{2}\longrightarrow \sqrt{2}x^{2}$ with respect to
the Euclidean level. $SO(2)_{STD.\left( \Pi _{2}\right) }$ is a 1-parameter
abelian group generated by $S_{SR}^{3}|_{x^{2}\longrightarrow \sqrt{2}x^{2}}$%
.
\[
\]

{\bf IV- Gravitational interaction.}

{\bf Case a).} The 5-d. metric (\ref{grav5D}) is:
\begin{equation}
g_{AB,DR5}(x^{5})=diag\left( 1,-b_{1}^{2}(x^{5}),-b_{2}^{2}(x^{5}),-1,\pm
f(x^{5})\right)
\end{equation}
Therefore the validity for $\mu =1,2$ of the $\Upsilon $ hypothesis (\ref{HP}%
) ({\em not} satisfied for $\mu =0,3$) depends on the nature and the
functional form of the metric coefficients $b_{1}^{2}(x^{5})$ and $%
b_{2}^{2}(x^{5})$.\medskip

{\bf Case b)}. The 5-d. metric (\ref{grav5D}) reads:
\begin{gather}
g_{AB,DR5}(x^{5})=diag\left( \frac{1}{4}\left( 1+\frac{x^{5}}{x_{0}^{5}}%
\right) ^{2},-b_{1}^{2}(x^{5}),-b_{2}^{2}(x^{5}),-\frac{1}{4}\left( 1+\frac{%
x^{5}}{x_{0}^{5}}\right) ^{2},\pm f(x^{5})\right) .  \nonumber \\
\end{gather}

By making suitable assumptions on the functional form of the coefficients $%
b_{1}^{2}(x^{5})$ and $b_{2}^{2}(x^{5})$, it is possible in 11 cases (which
include {\em all} cases of physical and mathematical interest) to solve the
relevant Killing equations for the gravitational interaction [28].

\subsubsection{The 5-d. ''$\Upsilon $-violating'' ($\forall \protect\mu
=0,1,2,3$) metrics of gravitational interaction.}

The ''$\Upsilon $-violating'' gravitational metrics can be discussed by
exploiting a general treatment of such a case [28]. If the $\Upsilon $
hypothesis (\ref{HP}) is {\em not} satisfied for $\mu =0,3$ in the energy
range $x^{5}>x_{0}^{5}$ , then the metric coefficient $f(x^{5})$ obeys the
following equation:
\begin{equation}
f^{\prime }(x^{5})+\frac{2}{x^{5}+x_{0}^{5}}f(x^{5})-\frac{2\complement }{%
x^{5}+x_{0}^{5}}\left( f(x^{5})\right) ^{2}=0,\complement \in R.
\end{equation}
This ODE belongs to the {\em separable subclass of Bernoulli type} $\forall
\complement \in R$. Solving it one gets the following 5-d. metric:
\begin{gather}
g_{AB,DR5}(x^{5})=diag\left( \frac{1}{4}\left( 1+\frac{x^{5}}{x_{0}^{5}}%
\right) ^{2},-b_{1}^{2}(x^{5}),-b_{2}^{2}(x^{5}),\right.  \nonumber \\
\nonumber \\
\left. -\frac{1}{4}\left( 1+\frac{x^{5}}{x_{0}^{5}}\right) ^{2},\pm \left(
\gamma \left( 1+\frac{x^{5}}{x_{0}^{5}}\right) ^{2}+\complement \right)
^{-1}\right) ,
\end{gather}
where in general parameters $\gamma $ and $\complement $ are real and
positive ({\em not both zero}).\bigskip

The case when the hypothesis $\Upsilon $ (\ref{HP}) is {\em not} satisfied
for $\mu =1$ and/or $2$ corresponds to metric coefficients $b_{1}^{2}(x^{5})$%
, $b_{2}^{2}(x^{5})$ and $f(x^{5})$ satisfying the following ODE (ESC off)
\begin{gather}
-\left( b_{i}^{\prime }(x^{5})\right) ^{2}+b_{i}(x^{5})b_{i}^{\prime \prime
}(x^{5})-\frac{1}{2}b_{i}(x^{5})b_{i}^{\prime }(x^{5})f^{\prime
}(x^{5})(f(x^{5}))^{-1}-c_{i}f(x^{5})=0,  \nonumber \\
\nonumber \\
c_{i}\in R,i=1\text{ and/or }2,  \label{135}
\end{gather}
whose solution in terms of $f(x^{5})$ is:
\begin{eqnarray}
f(x^{5}) &=&\frac{\left( b_{i}^{\prime }(x^{5})\right) ^{2}}{%
d_{i}b_{i}^{2}(x^{5})-c_{i}}\Leftrightarrow  \nonumber \\
&&  \nonumber \\
&\Leftrightarrow &d_{i}b_{i}^{2}(x^{5})f(x^{5})-\left( b_{i}^{\prime
}(x^{5})\right) ^{2}-c_{i}f(x^{5})=0,i=1\text{ and/or }2,  \label{136}
\end{eqnarray}
\begin{equation}
d_{i}\in R^{+},c_{i}\in R^{-}\text{{\footnotesize (not both zero)}.}
\end{equation}
The non-linear ODE (\ref{136}) can be solved in {\em all} possible cases:

1) $d_{i}\in R_{0}^{+},c_{i}\in R_{0}^{-}$;

2) $d_{i}=0,c_{i}\in R_{0}^{-}$;

3) $c_{i}=0,d_{i}\in R_{0}^{+}$,

(even in the limit case of $b_{i}$ constant). We refer the interested reader
to Ref. [28].
\[
\]

The above general formalism allows one to deal with \ the 5-d. metrics of
DR5 for the gravitational interaction which violate $\Upsilon $ $\forall \mu
=0,1,2,3$ in the energy ranges $0<x^{5}\leqslant x_{0(grav)}^{5}$ and $%
x^{5}>x_{0(grav)}^{5}$.

In the first case ($0<x^{5}\leqslant x_{0(grav)}^{5}$), the functional form
of $f(x^{5})$ is {\em undetermined}, since in general it must {\em only}
satisfy the condition $f>0$ $\forall x^{5}\in R_{0}^{+}$.

In the second case ($x^{5}>x_{0(grav)}^{5}$), one gets {\em 27 expressions
for the 5-d. gravitational metrics}. Their general functional form is:
\begin{gather}
g_{AB,DR5(grav.)}(x^{5})=diag\left( \frac{1}{4}\left( 1+\frac{x^{5}}{%
x_{0}^{5}}\right) ^{2},-b_{1}^{2}(x^{5}),-b_{2}^{2}(x^{5}),\right.  \nonumber
\\
\nonumber \\
\left. -\frac{1}{4}\left( 1+\frac{x^{5}}{x_{0}^{5}}\right) ^{2},\pm \left(
\gamma \left( 1+\frac{x^{5}}{x_{0}^{5}}\right) ^{2}+\complement \right)
^{-1}\right) ,
\end{gather}
where parameters $\gamma $ and $\complement $ are real and positive ({\em %
not both zero}). The explicit expressions of $b_{1}^{2}(x^{5})$ and $%
b_{2}^{2}(x^{5})$ (and therefore of the 27 gravitational metrics) can be
found in Ref. [28]. All such metrics satisfy the $\Upsilon ${\em -violating
equation} (\ref{135}) that, for fixed $\mu \in \left\{ 0,1,2,3\right\} $,
reads
\begin{equation}
-\left( b_{\mu }^{\prime }(x^{5})\right) ^{2}+b_{\mu }(x^{5})b_{\mu
}^{\prime \prime }(x^{5})-\frac{1}{2}b_{\mu }(x^{5})b_{\mu }^{\prime
}(x^{5})f^{\prime }(x^{5})(f(x^{5}))^{-1}-c_{\mu }f(x^{5})=0,  \label{139}
\end{equation}
with the following ranges and conditions:
\begin{eqnarray}
f(x^{5}) &\in &R_{0}^{+}\ \forall x^{5}\in R_{0}^{+} \\
&&  \nonumber \\
b_{\mu }(x^{5}) &\in &R_{0}\ \forall x^{5}\in R_{0}^{+}\Leftrightarrow
b_{\mu }^{2}(x^{5})\in R_{0}^{+}\ \forall x^{5}\in R_{0}^{+} \\
&&  \nonumber \\
c_{\mu } &\in &R.
\end{eqnarray}
The solution of non-linear ODE (\ref{139}) in terms of $f(x^{5})$ is:
\begin{equation}
f(x^{5})=\frac{\left( b_{\mu }^{\prime }(x^{5})\right) ^{2}}{d_{\left( \mu
\right) }b_{\mu }^{2}(x^{5})-c_{\mu }}.
\end{equation}

In correspondence to the above-mentioned 27 different gravitational metrics,
one gets {\em 27 systems of 15 Killing non-linearly coupled PDEs}, which it
is very difficult to solve explicitly.

Analogous results hold true in the $\Upsilon $ -violating case for the
gravitational interaction in the energy range $0<x^{5}\leqslant x_{0}^{5}$ ,
namely the same functional forms of $b_{i}^{2}(x^{5}),i=1,2$ as before are
obtained, but one has to put $b_{0}^{2}(x^{5})=b_{3}^{2}(x^{5})=1$ and to
leave $f(x^{5})$ {\em undetermined} (but strictly positive $\forall x^{5}\in
R_{0}^{+}$). One gets therefore {\em ''}$f(x^{5})${\em -dependent''}, i.e.
in general {\em ''functionally parametrized''}, metrics.We refer the
interested reader to Ref. [28] for a deeper discussion of this case.

\section{\protect\bigskip Five-dimensional geodesics}

As a last topic in DR5, let us consider the geodesics in the {\em sui generis%
} five-dimensional Riemann manifold $\Re _{5}$ , in order to clarify their
possible physical meaning.

The geodesic equations are
\begin{equation}
\frac{d^{2}x^{A}}{d\tau ^{2}}+\Gamma _{BC}^{A}\frac{dx^{B}}{d\tau }\frac{%
dx^{C}}{d\tau }=0.  \label{GEO}
\end{equation}

Let us here confine ourselves to find solutions to this equation in the {\em %
Power Ansatz} for the metric coefficients (see case {\bf ii)} of Sect. 4).
In this case Eq.s (\ref{GEO}) explicitly read
\begin{equation}
\left\{
\begin{array}{l}
\dfrac{d^{2}t}{d\tau ^{2}}+\dfrac{q_{0}}{E}\dfrac{dt}{d\tau }\dfrac{dE}{%
d\tau }=0; \\
\\
\dfrac{d^{2}x}{d\tau ^{2}}+\dfrac{q_{1}}{E}\dfrac{dx}{d\tau }\dfrac{dE}{%
d\tau }=0; \\
\\
\dfrac{d^{2}y}{d\tau ^{2}}+\dfrac{q_{2}}{E}\dfrac{dy}{d\tau }\dfrac{dE}{%
d\tau }=0; \\
\\
\dfrac{d^{2}z}{d\tau ^{2}}+\dfrac{q_{3}}{E}\dfrac{dz}{d\tau }\dfrac{dE}{%
d\tau }=0; \\
\\
\\
\dfrac{d^{2}E}{d\tau ^{2}}+\dfrac{r}{2E}\left( \dfrac{dE}{d\tau }\right)
^{2}-\dfrac{1}{2E^{r+1}}\left[ q_{0}\left( \dfrac{E}{E_{0}}\right)
^{q_{0}}\left( \dfrac{dt}{d\tau }\right) ^{2}-\right. \\
\\
\left. -q_{1}\left( \dfrac{E}{E_{0}}\right) ^{q_{1}}\left( \dfrac{dx}{d\tau }%
\right) ^{2}-q_{2}\left( \dfrac{E}{E_{0}}\right) ^{q_{2}}\left( \dfrac{dy}{%
d\tau }\right) ^{2}-q_{3}\left( \dfrac{E}{E_{0}}\right) ^{q_{3}}\left(
\dfrac{dz}{d\tau }\right) ^{2}\right] =0.
\end{array}
\right.  \label{GEO2}
\end{equation}
The complete solutions of Eq.s (\ref{GEO2}) for {\em all} classes (I)-(XII)
of Sect. 4 can be found in Ref.s [28] and [31].

Here we shall confine ourselves to consider the solution of Eq.s (\ref{GEO2}%
) for the metric class (VIII) (${\bf \alpha }_{VIII}=(0,0,0,0,r)$ ),
corresponding to a four-dimensional Minkowski space-time with {\em %
undetermined} energy exponent (which represents, in our framework, the
electromagnetic interaction: see Ref. [12] for the phenomenological aspects
of this metric). Indeed, the solution of (\ref{GEO2}) reads, in this case
([18],[28]):
\begin{equation}
t=\frac{2}{C_{1}(2+r)}E^{\frac{2+r}{2}}+C_{2},  \label{145}
\end{equation}
where $C_{1}$ , $C_{2}$ are integration constants. Putting $C_{1}=C$ , $%
C_{2}=0$ Eq. (\ref{145}) becomes
\begin{equation}
E=C\frac{2+r}{2}t^{\frac{2}{2+r}},
\end{equation}
whence, for $r=-4$:
\begin{equation}
Et=-C.  \label{147}
\end{equation}
By assuming $C=-\hslash $,{\em \ Eq. (\ref{147}) takes a form which reminds
\ the quantum-mechanical, Heisenberg uncertainty relation for time and
energy.} Otherwise stated, we can say that {\em the geodesics in a
five-dimensional space-time, embedding a standard four-dimensional Minkowski
space, correspond to trajectories of minimal time-energy uncertainty. }This
result (first derived in Ref. [18], and rigorously analyzed and generalized
in Ref. [28]), although preliminary, seemingly indicates that the
five-dimensional scheme of DR5 may play a role toward understanding certain
aspects of quantum mechanics {\em in purely classical (geometrical) terms}.
It agrees with Wesson's results on the connection between {\em Heisenberg's
principle} and {\em Kaluza-Klein theory in the STM model} ([29], [30]).

\section{Conclusions and perspectives}

The DR5 formalism lends itself to a number of possible, future developments.

These include e.g. solving the general Einstein equations with a n{\em %
on-zero cosmological constant}, $\Lambda \neq 0$. Further improvements of
the predictive power of the theory may come from the explicit introduction
of a space-time-coordinate dependence in the fifth metric coefficient $f$
and/or in the cosmological constant $\Lambda $, i.e. assuming
\[
f=f(E,x)\text{ \ and/or \ }\Lambda =\Lambda (E,x).
\]
As it is easily seen, this amounts to taking into account also the presence
of matter in our scheme. Clearly, solving the five-dimensional Einstein
equations in such a case is expected to be a quite formidable task.

A further topic deserving investigation is that of the {\em five-dimensional
action}. The {\em Einstein-Hilbert action} in $\Re $ reads, in this case:
\begin{equation}
S=-\frac{1}{16\pi \tilde{G}}\int d^{5}x\sqrt{\pm \tilde{g}}R,  \label{44}
\end{equation}
where $\tilde{g}=\det g_{DR5}$ , $\tilde{G}$ is the gravitational constant
and the double sign in the square root accords to that in front of $f$.
Among the problems concerning $S$, let us quote its physical meaning (as
well as that of $\tilde{G}$) and the meaning of those energy values $\bar{E}$
such that $S(\bar{E})=0$ (due to a possible {\em degeneracy} of the metric).

The {\em Killing symmetries} of DR5 deserve further investigation on many
respects. Let us quote, for instance:

{\bf 1} - the Lie nature of the {\em infinitesimal symmetries} derived;

{\bf 2} - the passage from the {\em infinitesimal} level to the {\em finite}
one;

{\bf 3} - and, last but not least, the {\em physical meaning} of the
symmetries obtained. As far as this last point is concerned, the results
obtained seemingly show {\em an invariance of physical laws under non-linear
coordinate transformations (in particular in time and energy).\bigskip }

Besides the above {\em ''classical''} problems, there are also what we may
call the possible {\em ''quantum''} aspects of the formalism. They are
related to the fact that actually, in most systems of physical interest at a
microscopical level, energy is {\em quantized}. {\em How does energy
quantization match in this scheme? How to account for energy jumps within an
apparently completely classical framework?\bigskip }

{\bf Acknowledgements - }It is a pleasure and a duty to warmly thank M.
Francaviglia, who coauthored the DR5 theory with F.C. and R.M., for
continuous interest and encouragement.\pagebreak

\bigskip

\text{{\bf References}}

[1] For the early attempts to geometrical unified theories based on higher
dimensions, see e.g. M.A.Tonnelat: {\it Les Theories Unitaires de
l'Electromagnetisme et de la Gravitation }(Gautier-Villars, Paris,1965), and
references quoted therein.

[2] Th. Kaluza: {\it Preuss. Akad. Wiss. Phys. Math}. {\bf K1}, 966 (1921).

[3] O.Klein: {\it Z. Phys.} {\bf 37}, 875 (1926).

[4] P.Jordan: {\it Z.\ Phys}. {\bf 157}, 112 (1959), and references quoted
therein.

[5] Y.Thiry: {\it C.R. Acad. Sci.(Paris) }{\bf 226}, 216 (1948).

[6] R.L.Ingraham: {\it Nuovo Cim}. {\bf 9}, 87 (1952).

[7] J.Podolanski: {\it Proc. Roy. Soc}. {\bf 201, }234 (1950).

[8] See e.g. {\it Modern Kaluza-Klein Theories,} T. Appelquist, A. Chodos
and P.G.O. Freund eds. (Addison-Wesley, 1987); J.M.Overduin and P.S.Wesson:
{\it Phys. Rept.} {\bf 283}, 303 (1997), and references quoted therein.

[9] See e.g. {\it Supergravity in Diverse Dimensions}, A.Salam and E.Sezgin
eds. (North-Holland \& World Scientific, 1989).

[10] See e.g. G.Yu. Bogoslovsky:{\it \ Fortsch. Phys}. {\bf 42}, 2 (1994);
R.M. Santilli: {\it Elements of Hadronic Mechanics}, vol.I-III
(Naukova-Dumka Pub., Kiev, 1994).

[11] For a review of Finsler's generalization of Riemannian spaces, see e.g.
M.Matsumoto: {\it Foundation of Finsler Geometry and Special Finsler Spaces }%
(Kaiseisha Otsu, 1986), and references quoted therein.

[12] F.Cardone and R.Mignani: {\it Energy and Geometry - An Introduction to
Deformed Special Relativity }(World Scientific, Singapore, 2004), and Ref.s
therein.

[13] S.H.Aronson, G.J.Bock, H.-Y.Chang and E.Fishbach:{}{\ {\it Phys. Rev.
Lett.} {\bf 48}, 1306 (1982); {\it Phys. Rev. }D {\bf 28}, 495 (1983); }%
N.Grossman et al.:{\it \ Phys. Rev. Lett.} {\bf 59}, 18 (1987).

[14] For experimental as well as theoretical reviews, see e.g. B.L{%
\"{o}rstad: {\it ''Correlations and Multiparticle Production (CAMP)''},}{%
eds. M. Pluenner, S. Raha and R. M. Weiner (World Scientific, Singapore,
1991); D.H. Boal, C.K. Gelbke and B. K. Jennings: Rev. Mod. Phys. {\bf 62},
553 (1990); and references therein.}

[15] For reviews on both experimental and theoretical aspects of
superluminal photon tunneling, see e.g. G.Nimtz and W.Heimann:{\it \ Progr.
Quantum Electr.} {\bf 21}, 81 (1997); R.Y.Chiao and A.M.Steinberg : {\it ''}%
Tunneling Times and Superluminality{\it ''}, in {\it Progress in Optics},
E.Wolf ed., {\bf 37}, 346 (Elsevier Science, 1997); V.S.Olkovsky and
A.Agresti: in {\it Tunneling and its Implications, } D.Mugnai, A.Ranfagni
and L.S.Schulman eds.(World Sci., Singapore, 1997), p. 327.

[16] C.O.Alley: {\it ''}Relativity and Clocks'', in {\it Proc.of the 33rd
Annual Symposium on Frequency Control,} Elec.Ind.Ass., Washington, D.C.
(1979) ; ''Proper time experiments in gravitational fields with atomic
clocks, aircraft, and laser light pulses'' in {\it Quantum optics,
experimental gravity, and measurement theory,} P.Meystre and M.O.Scully eds.
(Plenum Press, 1983), p. 363.

[17] B.Finzi: {\it ``}Relativit\'{a} Generale e Teorie Unitarie'' (''General
Relativity and Unified Theories''){, }{in {\it Cinquant'anni di
Relativit\'{a} (Fifty Years of Relativity)}, M.Pantaleo ed. (Giunti,
Firenze, 1955), pages 194 and 204.}

[18] F.Cardone, M.Francaviglia and R.Mignani: {\it Gen. Rel. Grav.} A{\bf 30}%
, 1619 (1998); {\it Found. Phys. Lett.} {\bf 12}, 281 (1999); ibidem, {\bf 12%
}, 347 (1999); {\it Gen. Rel. Grav.} {\bf 31}, 1049 (1999).

[19] F. Cardone, A. Marrani and R. Mignani: {\it Found. Phys.} {\bf 34}, 617
(2004), {\tt hep-th/0505088}; ibidem {\bf 34}, 1155 (2004), {\tt %
hep-th/0505105}; ibidem {\bf 34}, 1407 (2004), {\tt hep-th/0505116}.

[20] F. Cardone, R. Mignani and V.S. Olkhovsky: {\it Jour. Phys. I France}
{\bf 7}, 1211 (1997); {\it Modern Phys. Lett.} B {\bf 14}, 109 (2000);{\it \
Phys. Lett.} A {\bf 289}, 279 (2001).

[21] UA1 Collaboration {\it : Phys.Lett.} {\bf B 226}, 410 (1989).

[22] See L.\ Sklar: {\it Space, Time, and Spacetime }(Univ. of Calif. Press,
Berkeley, 1976), chapt. IV.

[23] P.A.M.Dirac:{\it \ Proc. R. Soc. (London)} {\bf A333}, 403 (1973); {\it %
ibidem}, {\bf A338}, 439 (1974).

[24] F.Hoyle and J.V.Narlikar: {\it Action at a Distance in Physics and
Cosmology} (Freeman, N.Y., 1974).

[25] V.Canuto, P.J.Adams, S.-H.Hsieh and E.Tsiang: {\it Phys. Rev.} {\bf D16}%
, 1643 (1977); V.Canuto, S.-H.Hsieh and P.J.Adams: {\it Phys. Rev. Lett.}
{\bf 39}, 429 (1977).

[26] P.S.Wesson : {\it Space-Time-Matter --- Modern Kaluza-Klein Theory}
(World Scientific, Singapore, 1999), and references therein.

[27] T.Fukui : {\it Gen. Rel. Grav}. {\bf 20}, 1037 (1988); {\it ibidem},%
{\bf \ 24}, 389 (1992).

[28] A. Marrani: {\it ''Aspetti matematici della Relativit\`{a} Deformata in
5 dimensioni''} ({\it ''Mathematical aspects of Deformed Relativity in 5
dimensions''}) (Ph.D. thesis, Dept. of Physics ''E.Amaldi'', Univ. ''Roma
Tre'', Rome 2004) (in Italian).

[29] P.S.Wesson and J.Ponce de Leon: {\it Astron. Astrophys.} {\bf 294}, 1
(1995).

[30] P.S.Wesson: {\it Mod. Phys. Lett}. {\bf A 10}, 15 (1995); {\it Gen.
Rel. Grav.} {\bf 36}, 451 (2004).

[31]{\bf \ }A. Marrani: {\it ''Simmetrie di Killing di Spazi di Minkowski
generalizzati''} ({\it ''Killing Symmetries of Generalized Minkowski Spaces''%
}) (Laurea Thesis), Rome, October 2001 (in Italian).

[32] F. Cardone, A. Marrani, and R. Mignani, {\it Found. Phys.} {\it Lett.}
{\bf 16}, 163 (2003), {\tt hep-th/0505032}.\bigskip

[33] F. Cardone, A. Marrani, and R. Mignani, ''The electron mass from
Deformed Special Relativity'', {\it Electromagnetic Phenomena} V.{\bf 3}, n.%
{\bf 1}(9) (2003), special number dedicated to Dirac's centenary, {\tt %
hep-th/0505134}.

[34] F. Cardone, A. Marrani, and R. Mignani, ''A geometrical meaning to the
electron mass from breakdown of Lorentz invariance'', published on the
monography {\it ''What is electron?''}, Apeiron (2005), {\tt hep-th/0505137}.

\end{document}